\documentclass[12pt,preprint]{aastex}
\usepackage{graphicx}
\usepackage{longtable}
\newcommand{\dg}{\ifmmode{\rm ^{\circ}}\else{$^{\circ}$}\fi}
\newcommand{\mic}{\ifmmode{\rm \mu m}\else{$\mu$m}\fi}

\newcommand{\iras}{{\em IRAS}}

\newcommand{\spi}{{\em Spitzer}}

\shorttitle{MIPSGAL Rings and Disks}
\shortauthors{Mizuno et al.}

\begin{document}

\title{A Catalog of MIPSGAL Disk and Ring Sources}

\author{ D. R. Mizuno\altaffilmark{1,7},  
    K. E. Kraemer\altaffilmark{2},
    N. Flagey\altaffilmark{3}, 
    N. Billot\altaffilmark{4}, 
    S. Shenoy\altaffilmark{5},
    R. Paladini\altaffilmark{3},
    E. Ryan\altaffilmark{6},
    A. Noriega-Crespo\altaffilmark{3}, 
    S. J. Carey\altaffilmark{3}
}

\altaffiltext{1}{Institute for Scientific Research, Boston College, 140 Commonwealth Ave., Chestnut Hill, MA, 02467-3862 USA}
\altaffiltext{2}{Air Force Research Laboratory, AFRL/RVBYB, 29 Randolph Road, Hanscom AFB, MA 10731 USA}
\altaffiltext{3}{Spitzer Science Center, MS 220-6, California Institute of Technology, Pasadena, CA 91125 USA}
\altaffiltext{4}{Infrared Processing and Analysis Center, MS 100-22, California Institute of Technology, Pasadena, CA 91125 USA}
\altaffiltext{5}{Ames Research Center, MS 245-6, Moffett Field, CA 94035 USA}
\altaffiltext{6}{University of Minnesota, Department of Astronomy, 116 Church St. S.E., Minneapolis, MN 55455 USA}
\altaffiltext{7}{afrl.rvb.pa@hanscom.af.mil}

\begin{abstract}

We present a catalog of 416 extended, resolved, disk- and ring-like objects as detected in the 
MIPSGAL 24 {\micron} survey of the Galactic plane. This catalog is the result of a search
in the MIPSGAL image data for generally circularly symmetric, extended ``bubbles'' without
prior knowledge or expectation of their physical nature.
Most of the objects have no extended counterpart at 8 or 70 \micron, with less
than 20\% detections at each wavelength. For the 54 objects with central point sources, the sources 
are nearly always seen in all IRAC bands. 
About 70 objects (16\%) have been previously identified, with another
35 listed as IRAS sources. Among the identified objects, 
those with central sources are mostly listed as emission-line stars, but with
other source types including supernova remnants, luminous blue variables, and planetary nebulae.
The 57 identified objects (of 362) without central sources are nearly all PNe ($\sim$90\%),
which suggests that a large fraction of the 300$+$ unidentified objects in this category are also
PNe. These identifications suggest that this is primarily a catalog of evolved stars. 
Also included in the catalog are two filamentary objects that are almost certainly
supernova remnants, and ten unusual compact extended
objects discovered in the search. Two of these show remarkable spiral structure 
at both 8 and 24 \micron. These are likely background galaxies previously hidden by the 
intervening Galactic plane. 

\end{abstract}


\section{Introduction\label{sec.intro}}

When a star evolves up the asymptotic giant branch (AGB), its atmosphere 
expands and cools. The ejected gas can condense into dust grains within the 
circumstellar shell, which may become quite bright in the infrared (IR), 
while at the same time often becoming sufficiently optically thick to hide 
the star itself in the optical. As the star continues to evolve and shed mass, 
it becomes a post-AGB object, and eventually a planetary nebula (PN) or 
supernova (SN). These objects at the end stages of the stellar lifecycle are 
responsible for creating of most of the dust in the Universe 
\citep[e.g.][]{gehrz89} and injecting it 
into the interstellar medium, where it can then form into new stars and 
planets. Because of the dust content of the circumstellar shells around AGB 
stars and in the ejecta of PNe and SNe, the IR is an ideal wavelength regime in
which to identify new evolved objects. This is particularly true in the 
Galactic plane where high extinction in the optical limits searches to
nearby sources. Previous IR surveys such as the Air Force Geophysics Laboratory
(AFGL; Walker \& Price 1975) and {\em Infrared Astronomy
Satellite} (\iras; Olnon et al. 1986) surveys found
hundreds of AGB stars due to their dust emission, although they generally 
could not resolve the circumstellar structures around the stars.

Identification of new evolved star candidates and their circumstellar shells 
has recently been facilitated with the advent of high resolution mid-infrared 
imaging surveys.  While full characterization of the progenitors may require 
spectroscopy, IR imaging is often the easiest and most telescope-time 
efficient way to identify evolved stars, particularly along lines of sight 
with high extinction. Specifically, the \spi\ Legacy surveys of 
the inner Galaxy, MIPSGAL \citep{carey09} with the Multiband Imaging 
Photometer for \spi (MIPS) instrument at 24 
and 70 \micron \citep{mips}, and the Galactic Legacy
Infrared Mid-Plane Survey Extraordinaire \citep[GLIMPSE;][]{benjamin03}, with the Infrared
Array Camera (IRAC) at 3.6--8.0 \micron \citep{irac}, 
provide high resolution (1.8 to 6\arcsec) imaging of the majority of the 
inner Galactic disk, which should include a large fraction of the evolved stars in 
the Milky Way. Ring and bubble structures associated with evolved 
stars, PNe, and supernova remnants (SNRs) are readily identifiable in these
datasets. Indeed, several studies of rings found in the GLIMPSE archive, i.e. 
at 3.6--8.0 \micron, have already been
made, such as \citep{phillips08} or \citep{ch06}, although these were
primarily from massive young stars and only a few SNRs. Here, we present a catalog
of ring and disk sources found in the MIPSGAL data at 24 \micron.

\section{Identification of Sources\label{sect.ids}}

The MIPSGAL 24 \micron\ mosaics (Mizuno et al. 2008; Carey et al. 2009) were 
searched by visual examination for 
candidate ``bubble'' objects. (Visual inspection is still the most reliable way to
detect sets of extended objects; no automated procedure yet exists to replace
the human eye.) 
The criteria for inclusion were: (1) 
Generally round shape with a hard-edged boundary; with either (2) an approximately 
flat ``disk'' profile or (3) presence of a ring or partial ring, allowing some amount 
of regular or irregular structure or irregular shape. In particular we excluded 
round and extended but centrally
peaked objects: those of small angular size are either unresolved or barely resolved 
at the 6{\arcsec} resolution of the MIPSGAL survey
and thus have uncertain underlying morphology; larger, 
and likely resolved, examples of such objects have also been excluded 
from the candidate set somewhat arbitrarily. In addition to the round extended 
objects specifically sought, we were also attentive to any compact extended 
objects that had a common morphology; we identified and included an ensemble 
of objects that have a bipolar or ``two-lobed'' appearance.

A total of 416 objects were selected. The objects were 
grouped, by a visual assessment of their morphology, 
into four primary categories. The first is
objects with a detected central point source at 24 \micron. The objects 
without central point sources are separately categorized as rings, disks, 
or two-lobed. Each group is further divided into subgroups depending
on particular symmetry and regularity properties. Note that the groups
and subgroups are defined entirely by the visual appearance of the objects;
we make no a priori claim that these groupings represent underlying physical 
distinctions apart from the obvious morphology. 
In addition, two small additional categories are
included: filamentary objects (two items, both identified as SNRe) and a miscellaneous
category (ten objects) containing singular compact extended objects.

\begin{figure}
\epsscale{0.75}
\plotone{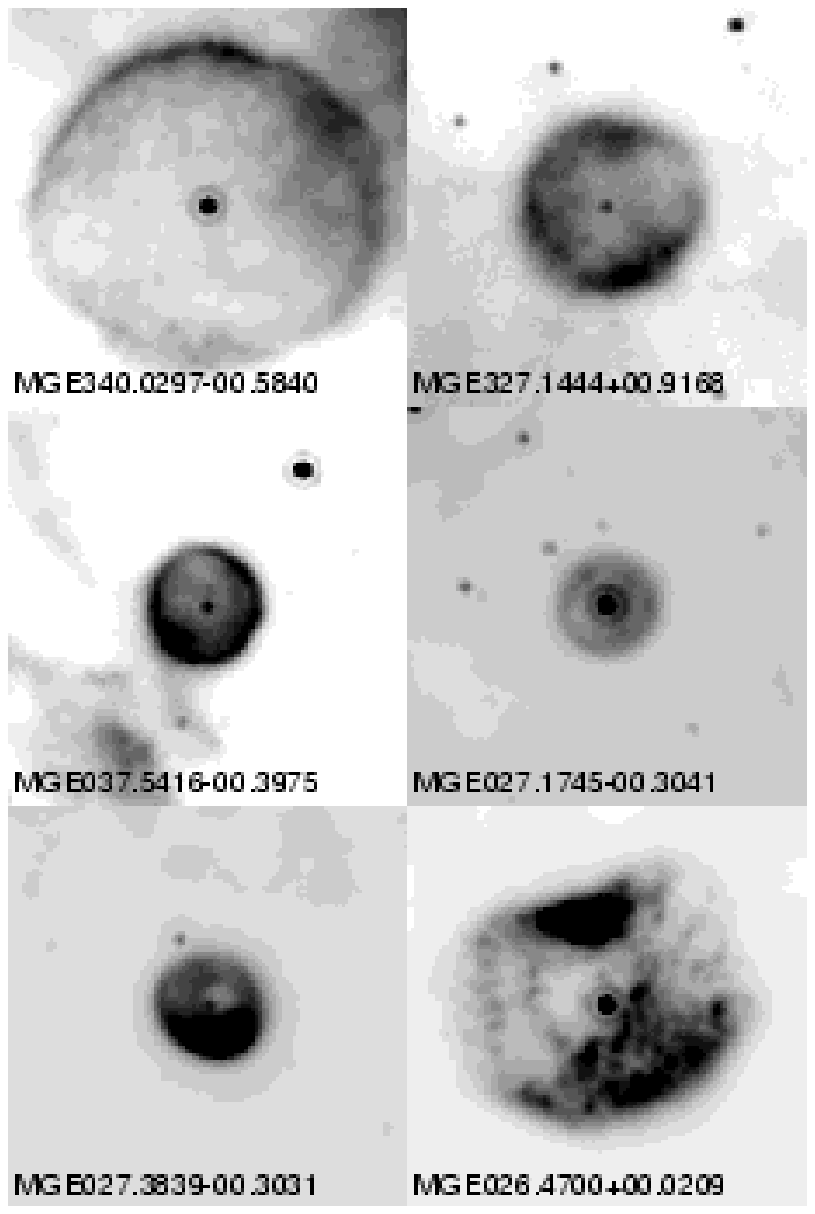}
\caption{Examples of objects with central sources. The bottom row are ``irregular'' examples. 
Each panel is about 3{\farcm}3 across.}
\label{central}
\end{figure}

\begin{figure}
\epsscale{0.75}
\plotone{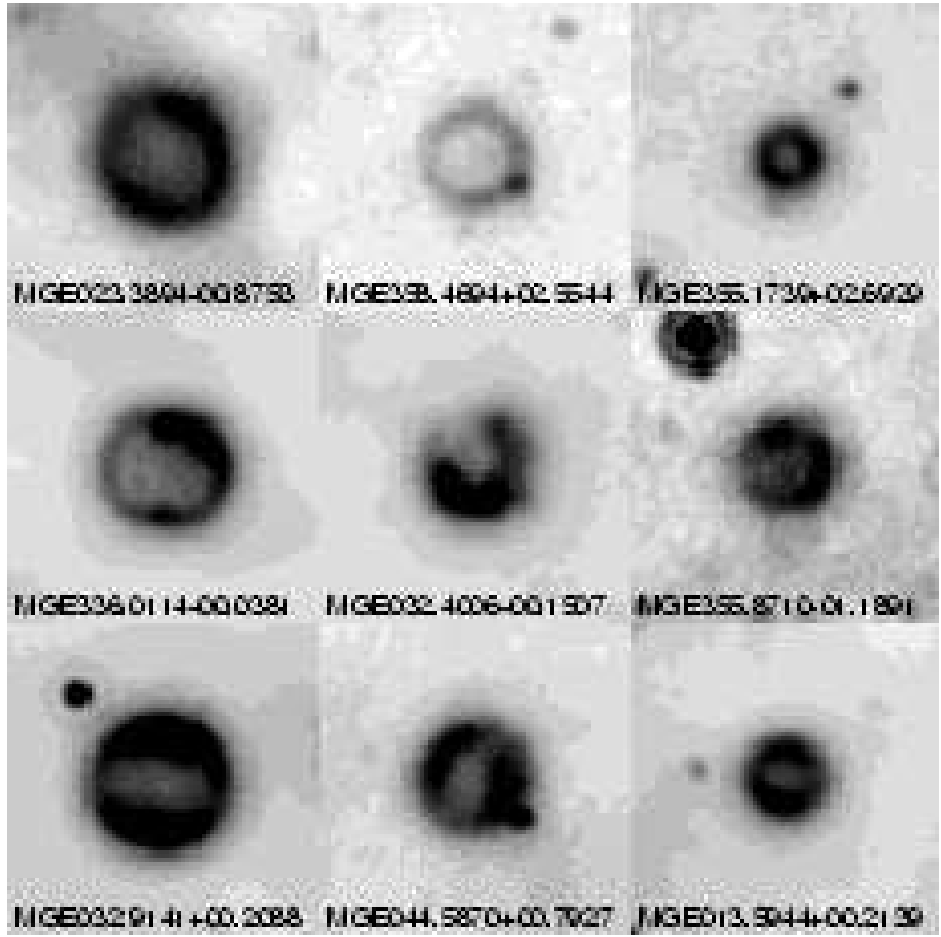}
\caption{Examples of ring objects. Top row: Rings. Middle: Irregular rings. Bottom: Bilaterally 
symmetric. Each panel is about 1{\farcm}5 across.}
\label{rings}
\end{figure}

\begin{figure}
\epsscale{0.75}
\plotone{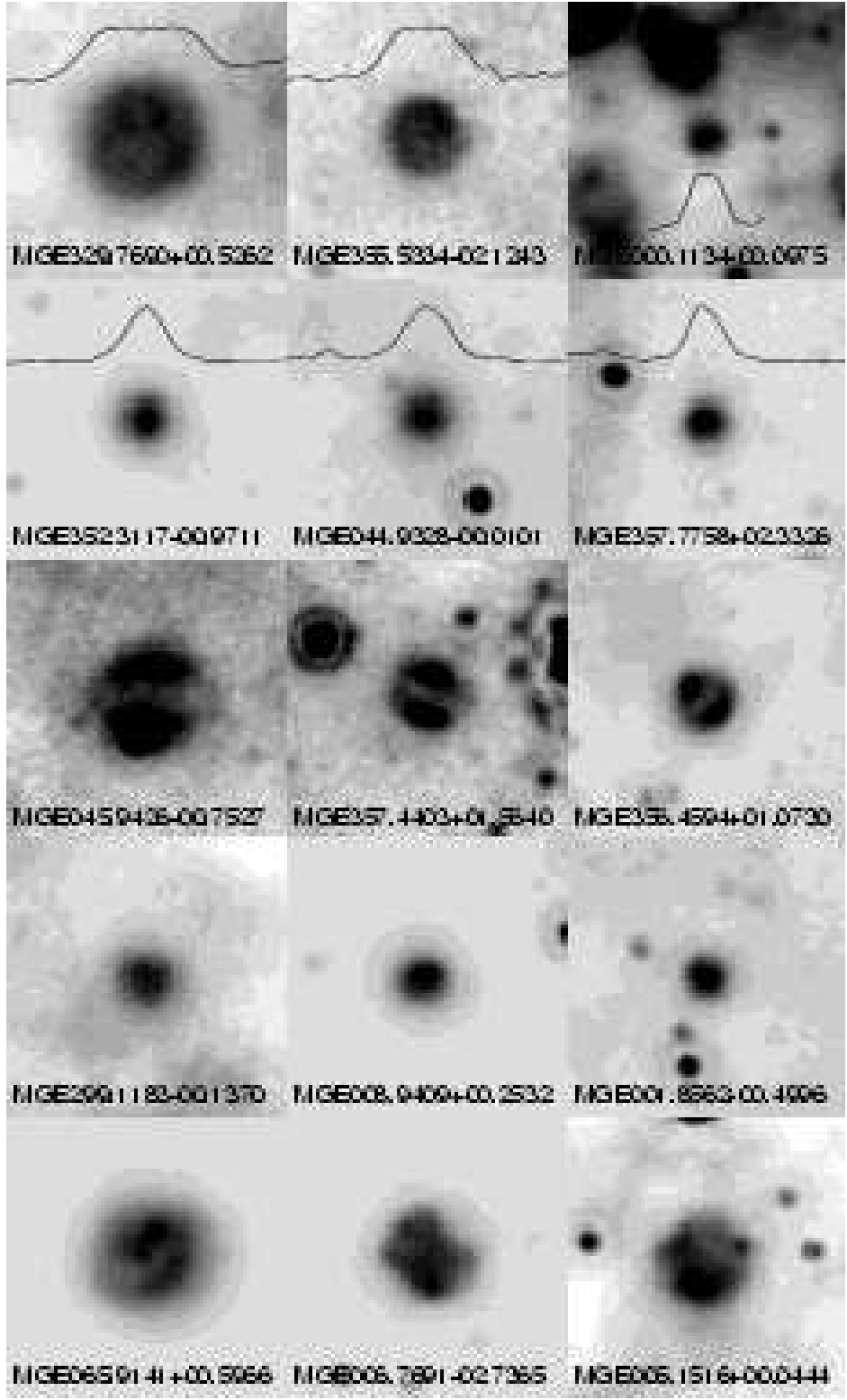}
\caption{Examples of disk objects. Top to bottom: flat, peaked, bilaterally symmetric, oblong, and irregular.
Each panel is about 1{\farcm}5 across. The flat and peaked examples also have horizontal slices included 
to show the distinction between these subgroups.}
\label{disks}
\end{figure}

\begin{figure}
\epsscale{0.75}
\plotone{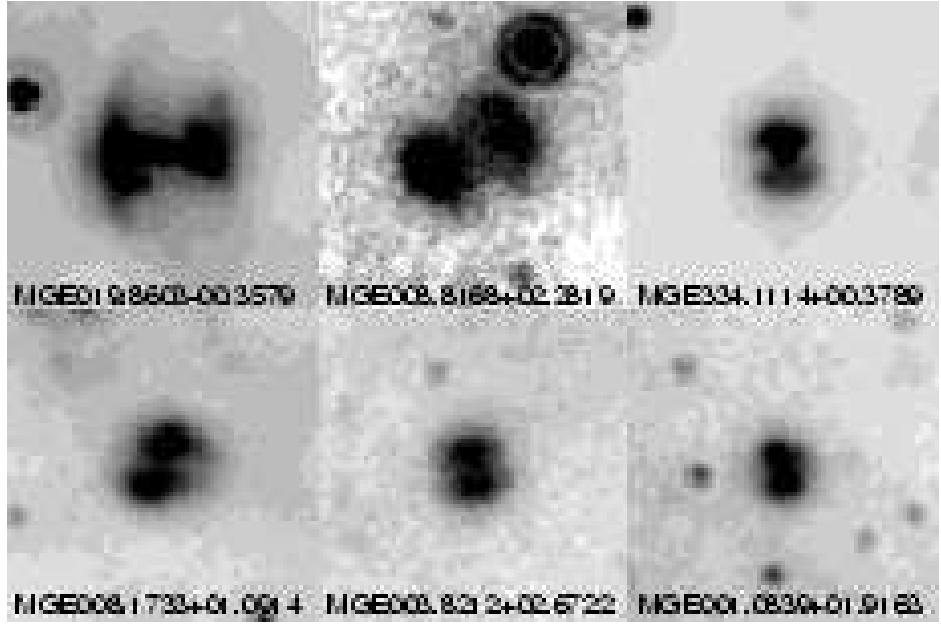}
\caption{Examples of two-lobed objects. Each panel is about 1{\farcm}5 across.}
\label{twolobed}
\end{figure}

\begin{figure}
\epsscale{0.75}
\plotone{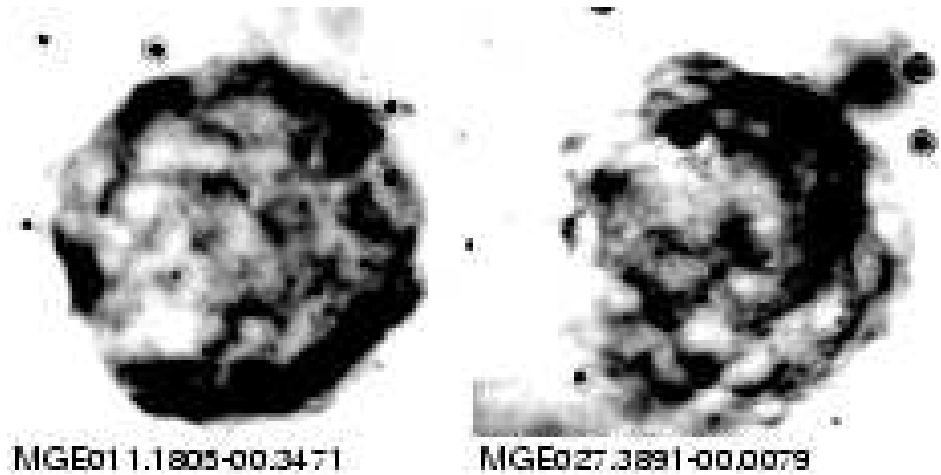}
\caption{Filamentary objects. Each panel is about 5{\farcm}6 across.}
\label{filamentary}
\end{figure}

\begin{description}

\item{Objects with central point sources (Group 1):} The general morphology is almost
invariably ringlike, which is not surprising as the visibility of a central
source implies an optically thin shell. Figure \ref{central} shows a few 
representative examples. The median angular size of these objects,
about 44{\arcsec}, is approximately twice that of the objects lacking a visible
24 {\micron} central source. Only three of objects lacking a central source 
are larger than 44{\arcsec} (in a six times more numerous ensemble) so this difference 
indicates that these are a distinct population of objects apart from the visibility of
the central source itself. {\em Regular} (1a) examples are, in gross features,
axially symmetric or bilaterally symmetric. {\em Irregular} (1b)
examples feature highly nonuniform or nonaxisymmetric ring brightness, 
or significant extended structure apart from the ring itself.

\item{Rings (without central sources) (Group 2):} These are divided into three 
subcategories. {\em Rings} (2a) are complete rings with some allowed unevenness in 
ring brightness or morphology but no other significant structure.
Included in this subcategory are both thin rings and the more common 
``thick ring'' object type whose appearance is as a flat disk with a central
depression. {\em Irregular rings} (2b) are objects featuring either a partial ring
or a complete ring but with large variations in brightness or thickness around the
ring. {\em Bilaterally symmetric rings} (2c) have an axis across which the
ring has either symmetrically enhanced brightness or a change in shape such 
that the ring typically acquires a ``D'' shape on either side. Figure \ref{rings}
shows examples of each of these three subcategories.

\item{Disks (Group 3):} These have five subcategories: {\em Flat} disks (3a) are axisymmetric, largely
featureless disks with an essentially flat profile; ``flat'' is here assessed using 
a radial profile calculated for each disk from the 24 {\micron} image data, 
consisting of an azimuthal average at each 
radius in (1{\farcs}25) pixel increments, and is defined 
as a less than $5\%$ drop from the peak brightness at the third pixel radius in the 
profiles, i.e. a region about a resolution element wide in solid angle. Some minor structure 
is allowed as there are no truly featureless, flat disks. {\em Peaked}
disks (2b) are axisymmetric but fail the flatness criterion in the radial profiles. 
Note that circularly symmetric but strongly peaked extended objects have generally been
excluded from consideration for this catalog, so this subgroup is just a partial sampling of such
objects. {\em Bilaterally symmetric}
disks (2c) have a symmetry axis with enhanced brightness on either side; typically 
these look like otherwise flat disks with a slot across the center. {\em Oblong} disks (2d) are
featureless but have a slightly elliptical or elongated shape rather than round. 
{\em Irregular} disks (2e) have either pronounced asymmetric structure or irregular
shape. Figure \ref{disks} shows examples of the disk subcategories. The flat and peaked
examples also show horizontal profiles to demonstrate the distinction between these subgroups.

\item{Two-lobed (Group 4):} Unlike the other categories, these objects are not 
primarily round. Rather, they consist of two small lobes typically separated by a
much fainter lane perpendicular to the emission lobes. Figure \ref{twolobed} shows a few
examples. One lobe is usually brighter
than the other, presumably due to the viewing geometry. While many of the bilaterally
symmetric rings and disks have a generally two-lobed appearance, this category contains objects
that have a specific two-lobed boundary and no underlying disk. 

\item{Filamentary (Group 5):} This category includes localized, bounded objects with a primarily 
filamentary appearance, and contains just two objects which are identified as SNRs
(Figure \ref{filamentary}).

\item{Miscellaneous (Group 6):} This category contains a small number of singular compact 
extended objects that were discovered in the search for the disks and rings. Figure \ref{odd}
shows eight of the ten; the remaining two are possible spiral galaxies and are addressed in the Appendix. 

\end{description}

We make no claims regarding completeness of the catalog as a whole or for any of the 
morphology types, although
an effort at a thorough visual search of all the MIPSGAL mosaics was made.
Detection depends not only on the complexity and background variation
of a given region of the 
mosaics, which are significant factors in a visual search, but also on the 
intrinsic properties of the sources themselves. In particular, low surface brightness
objects against backgrounds with steep gradients are likely to be missed, although we
have not quantified the detection problem. 

\subsection{Statistical overview}

Of the 416 total objects, we identified 54 objects with central sources, 112 rings, 
226 disks, 24 two-lobed objects, plus the two filamentary and 10 miscellaneous objects. 
Overall, there is slightly more than one object per square degree ($\sim$340 square degrees
in the MIPSGAL survey), with a higher density near the Galactic center ($\sim$2 per square 
degree for $|l| < 10{\dg}$) and lower away (0.85 per square degree for $l > 10{\dg}$ and 0.6 per
square degree for $l < 350{\dg}$). The sensitivity is approximately constant over the entire
MIPSGAL survey region.

Figure \ref{galhist} shows a histogram of the Galactic longitude of the 240 objects with a
latitude within 1{\dg} of the plane (we have data further from the plane only within
10{\dg} of the Galactic center), in 10{\dg} bins. There is a clear population 
enhancement in the Galactic center region, comprising about a quarter of the total, 
and a general falloff to larger longitudes
away from the center. There are also apparent enhancements at about 30{\dg} from the center, 
although the small numbers make it difficult to make definite claims about the 
distribution. The enhancement
at $l \sim $30{\dg} is about a 2$\sigma$ deviation from the general trending and is 
likely a real effect. The enhancement at $l \sim 330{\dg}$ is only a 1$\sigma$ deviation
from the mean trending, and is more ambiguous.

Figure \ref{dist} shows the 2-D distribution of the objects in Galactic coordinates, 
separated by object group. Here all the objects are displayed.
The dashed lines show the approximate boundaries of the MIPSGAL survey. The disks 
show a markedly higher population density at high latitude ($> 1\dg$ from the plane) near the 
Galactic center, but as we have no high latitude data elsewhere, interpreting this result is
problematic. Also, this increased high-latitude disk density is
likely to be at least in part a selection effect as the backgrounds at 24 {\micron} are much 
more quiescent away from the plane. The rings also show a modest high-latitude enhancement 
but otherwise seem not to be preferentially located in the plane. 
The central-source objects by contrast show a definite 
paucity in the high-latitude data but instead show an evident clustering in regions 
${\sim}30{\dg}$ from the Galactic center. 

Figure \ref{size} shows histograms of the angular sizes of the central-source objects, and
the rings and disks combined. The lower limit of about $10\arcsec$ is dictated by the search 
criterion that objects be either distinctly ringlike or show a flat profile; objects much
below $10\arcsec$ are either unresolved or barely resolved and thus are generally omitted from
consideration due to the typical peaked appearance of such sources. All the rings and disks are below
1{\arcmin} in diameter, and the vast majority are below $30\arcsec$, while the median of the central-source
objects is $44\arcsec$. The two-lobed objects are too few to provide a meaningful histogram 
but generally follow the rings and disks in size range.

\subsection{Flux measurements}

We performed aperture photometry to determine the 24 {\micron} fluxes of the objects. For this
purpose, we used both the original MIPSGAL mosaics and the point source-subtracted version 
of the mosaics (Shenoy et al., in preparation). For each 
object, an ON source radius is selected by examining both the 2-D images and 1-D vertical and
horizontal slices through the center of the object. The ON radius is chosen to minimally contain
the entire object. 

Similarly, the 2-D images and slices are inspected to determine the radii for the OFF annulus,
which are selected to match the apparent background surface brightness at the object, 
avoid nearby sources, and stay as close as possible to the ON radius. The
background intensity I$_{\scriptstyle BG}$ is determined as the median of the pixels in the OFF annulus. The background 
is subtracted from the ON circle, the ON pixels are summed, and the sum is
scaled by the pixel solid angle (for the MIPSGAL mosaics, $ \Omega_{pixel} = 3.67 \times 10^{-11}$ sr) to give the flux F$_{24}$,


\begin{displaymath}
\mbox{F}_{24} = \Omega_{\scriptstyle pixel} \sum_{\scriptsize i=1}^{\mbox{\scriptsize n}} (p_i - \mbox{I}_{\scriptstyle BG})
\end{displaymath}
\begin{equation}
 = \Omega_{\scriptstyle pixel} \left( \sum_{\scriptstyle i=1}^{\mbox{\scriptsize n}} p_i - \mbox{nI}_{\scriptstyle BG} \right)
\end{equation}
where the summation is over the n pixel values $p_{i}$ in the ON region.

For the central-source objects, the original mosaics are used to determine the fluxes because the
source subtraction is not reliable for the central sources (many are not strictly point-like). The
fluxes thus may be occasionally contaminated by other point sources occurring within the ON radius. For 
the remainder of the objects, and for all background measurements, the source-subtracted mosaics are used. 

The flux errors are determined using both the uncertainty map (a product of the mosaic 
generation in the MOPEX\footnote{The MOPEX software is available for download at 
http://ssc.spitzer.caltech.edu/postbcd/mopex.html} software package) and an
empirically measured background error estimate. The uncertainty map provides a per-pixel
error estimate ($\sigma_{\scriptstyle UNC,i}$) and is presumed to be uncorrelated across pixels. The measurement 
of the RMS over the OFF annulus pixels ($\sigma_{\scriptstyle OFF}$) reflects both pixel-to-pixel scatter and 
also variations that are correlated on some length scale due to true background fluctuations. To be conservative,
we assume that background fluctuations dominate the OFF annulus RMS, and therefore
interpret it to be the overall uncertainty in the measured background 
level I$_{\scriptstyle BG}$ assigned to the ON region, and so the error 
for nI$_{\scriptstyle BG}$ is n$\sigma_{\scriptstyle OFF}$.


With these assumptions, the error in the flux is expressed by
\begin{equation}
\sigma^{2}_{\mbox{\scriptsize F}_{24}} = \Omega^{2}_{\scriptstyle pixel}  \left( \sum_{\scriptstyle i=1}^{\mbox{\scriptsize n}}\sigma_{\scriptstyle UNC,i}^{2} + \mbox{n}^{2}\sigma_{\scriptstyle OFF}^{2} \right)
\end{equation}

Note that this is an upper limit because we are assuming a worst-case situation for the uncertainty in the 
background level measurement, which generally dominates the error expression, and also 
because $\sigma_{\scriptstyle OFF}$ is in part an empirical measure 
of some noise contributions that are already represented in the uncertainty map for the ON circle (e.g. Poisson 
noise from the background emission).

\section{IRAC and MIPS 70 \micron\ detections}

We searched the GLIMPSE images for counterparts in each of the shorter 
wavelength IRAC bands, as well as the MIPSGAL 70 {\micron} data 
(Paladini et al. in preparation). GLIMPSE data are available for 314 of 
the objects. Of these, roughly 14\% (44) of the objects are detected at 8 {\micron} 
with extended emission, 80\% (252) were definite non-detections, and
the remainder were ambiguous.  At 70 \micron, data are available for 368 objects,
with 19\% (60) detections and 46\% (149) non-detections (there is a considerable amount
of ambiguously associated emission at 70 \micron, overlapping with the objects but
with no obvious related morphology).
Figure \ref{pan9} shows examples of definite detections, non-detections, and ambiguous 
detections at 8 and 70 {\micron} for three disk objects. 

The detection fractions are somewhat higher for the objects with 24{\micron} central sources.
The central sources themselves are observed in all IRAC bands for 94\% of these objects.
(IRAC central sources are observed for about 11\% of the rings, disks, and two-lobed objects.)
For extended emission, at 8 {\micron} (50 objects with IRAC data), there are 24\% (12) 
detections and 56\% (28) non-detections. 
At 70 {\micron} (46 objects), there are 43\% (20) detections and 15\% (7) non-detections.

Figure \ref{fig.3c} shows three-color images of four objects using 3.6 and 8
 \micron\ IRAC data from the GLIMPSE survey (blue and green) and 24 \micron\
MIPSGAL data (red). The upper left panel shows an example where there is no
extended emission at 8 \micron\ (although there may be at 70 \micron). When
present, extended 8.0 \micron\ emission is typically either co-spatial (Fig.
\ref{fig.3c}, upper right) with the 24 \micron\ emission or interior to it
(Fig. \ref{fig.3c}, lower left). In only one instance, shown in the lower
right of Figure \ref{fig.3c}, does the 8.0 \micron\ emission appear to extend
beyond the 24 \micron\ structure. In this case, there is fainter emission at
24 \micron\ in the 8 \micron\ region, but the 24 \micron\ structure is
dominated by the bright emission that fills the central hole in the 8 \micron\
structure. The average size of the 24 \micron\ structures that also
have 8 \micron\ extended emission is $\sim30\arcsec$. Thus,
in most cases, the relative shapes and sizes of the 8 and 24 \micron\ emission
regions are readily apparent.

In the cases where the 8 and 24 \micron\ emission is co-spatial, the
emitting particles, presumably large molecules such as polycyclic aromatic
hydrocarbons (PAHs) and small grains, respectively, must be well-mixed where
the structures are well-resolved at both wavelengths (the diffraction limits
were $\sim2\arcsec$ and $\sim6\arcsec$ at 8 and 24 \micron, respectively), as
in the examples shown in Figure \ref{fig.3c}. For the smallest disks and
rings, it is harder to say, and these objects could, of course, turn out to be
more complex if observed at higher resolution.

For those instances where the emission structures are clearly not co-spatial,
the story is more complicated. Complex emission structures are commonly
seen in the visible from planetary nebulae, such as the well-known Ring
Nebula. That complexity,
however, is caused by a variety of ions of different excitation potentials.
Here, in a few instances where spectroscopy shows that
the 24 \micron\ emission is actually line emission from [O IV], there is no
detected 8 or 70 \micron\ emission, either point-like or extended (Flagey
et al. in preparation). Chu et al. (2009) observed 36 known Galactic
PNe with MIPS, comparing the 24 \micron\ emission with H$\alpha$ images. They
explain the spatial differences between their 24 \micron\ emission and the
H$\alpha$ emission as depending on the dust content and excitation/ionization
structures in a particular PN. Here too, the dust density distribution and the
spectral energy distribution of the central exciting source are probably also
combining to create the observed 8 and 24 \micron\ emission structures.
The different 'layers' of dust emission could
represent different episodes of massloss from the parent star while it was on
the AGB.  Radiative transfer models (e.g. Egan et al. 1988)
that account for the 2- or 3-dimensional dust density structures, the
spectral energy distribution of the exciting source (which will be quite
problematic for the objects where none such has been detected), etc., are
needed to fully describe these emission structures, as was done with
{\em MSX} data for similar objects (e.g. Egan et al. 2002, Clark et al. 2003),
but that is beyond the scope of this paper.

\section{SIMBAD Correlations}

To help determine the nature of these sources, we did a SIMBAD search around
each object\footnote{Given the updates that occasionally take place in 
SIMBAD we note that the searches were performed in October 2008.}. A search 
radius of 2{\arcmin} was used to ensure that 
objects with imprecise coordinates, particularly IRAS sources, were not 
missed. The results were then compared to the 24 {\micron} images to 
determine if a SIMBAD object correlated with our target or with another 
object in the field.  We found 105 objects out of the 428 (including 
filamentary and miscellaneous) with a counterpart in 
SIMBAD. In most cases where a match was judged to be real, the SIMBAD object
was within $10\arcsec$ from our source coordinates. If an IRAS source corresponds 
to a particular section of one of our sources, such as the brightest arc of a 
ring, this is noted in the table notes. For the three
supernova remnants, although the pulsar and other components might be closer
to the center of the structure we detect, i.e. nominally a closer match to the
coordinates, we give the association as the supernova remnant since we are
detecting the extended emission, not the pulsar. There are a few objects that
were associated with radio sources detected at a single wavelength. These
cases are sufficiently unusual that they may reflect chance associations 
despite being within a few arcseconds of our 24 {\micron} objects.

\subsection{Morphology and Source Type}

Since approximately three-quarters of our sources are unknown, we can only draw conclusions 
as to what they are by extension from the sources with known associations. As 
mentioned above, among the categories returned with the SIMBAD results is 
the object type. For objects that have been previously studied, a literature 
search may reveal further information about the object (for example, two 
luminous blue variables are identified simply as stars by SIMBAD). Here, 
we describe the kinds of previously identified objects in each of the four 
main categories. 

\paragraph{Objects with central sources:} Just under half (24/54) of the objects
with central stars have SIMBAD counterparts, associated with either the central
source or the extended emission.  Eight are associated with
emission-line stars, including two luminous blue variables and one post-AGB
star. Two others are planetary nebulae, one is a supernova remnant, and
two are identified as stars, with no further information available in the
literature beyond their spectral types (B9 and M2). The remaining 11 are identified
only as IRAS sources with no additional information available, except perhaps 
2MASS{\footnote{Most MGE sources have at least
one 2MASS source within 5{\arcsec}, often more than one, which is not surprising
given the density of sources in the Galactic plane. However, unless there is  
corresponding point source at IRAC wavelengths, a chance alignment cannot be 
ruled out.}} data.

\paragraph{Rings:} Of the 112 ring objects, 22 have associations in SIMBAD. Of 
these, over half (13) are PNe, distributed fairly evenly among the three ring
subgroups (given the small numbers involved). Six more are IRAS 
sources, two are radio wavelength objects, and one is a B9 star. As with 
the central star objects, these 9 have no additional references.

\paragraph{Disks:} The SIMBAD associations among the 226 disk objects are 
predominently planetary nebulae: 36 of the 42 objects with counterparts are 
PNe. The remainder include three otherwise anonymous IRAS sources, two radio 
wavelength objects, and one eclipsing binary.

\paragraph{Two-lobed:}  Only five of the thirty two-lobed objects
have SIMBAD counterparts. Two are PNe identified in the Macquerie/AngloAustralian 
Observatory/Strasbourg H$\alpha$ planetary nebula survey
(Parker et al. 2006; Miszalski et al. 2008). Two have an IRAS
association with no other information available.  The last is associated with 
a star in the young open cluster NGC 6383, and as with the small number of 
radio sources, may be a chance spatial coincidence.

\section{The catalog}

Table 1 is the catalog of all the objects. The table is divided into object group
and subgroups, and within each subgroup the objects are sorted by increasing Galactic
longitude.

\begin{description}

\item{Name:} Constructed on the Galactic longitude and latitude.

\item{J2000 Coordinates:} The centers of the objects were selected by eye and specified
as the nearest pixel in the MIPSGAL mosaics, so the precision for the coordinates is 
good to about 2{\arcsec}.

\item{Diameter:} The boundaries of the objects are evaluated by eye in the mosaics, and 
the diameter on the horizontal axis (for round objects) is specified as the nearest 
integral pixel span, converted to arcseconds and rounded to integral values.
For objects with irregular boundaries, the larger of the horizontal and vertical axes is
used.

\item{24 {\micron} flux:} As described in Section 2.2. The errors calculated from equation (2) 
are shown as percentages.  Errors greater or equal to 100\% are truncated to 99\%.

\item{Detection flags:} These are determined by visual inspection of the 8 {\micron} GLIMPSE 
images and 70 {\micron} MIPSGAL images. For point sources, an affirmative result (``Y'') 
required a point source at the geometric center of the object and be free of confusion. An ambiguous 
result (``?'') is indicated if a source is slightly off-center or there is confusion present.
A negative result (``N'') means clearly no point source present. A dash means no data are available.
For extended emission, an affirmative result required a morphology similar to the 24 {\micron} 
morphology (or at least concentric), an ambiguous result means some local extended emission 
present but of uncertain correspondence
to the 24 {\micron} emission, and a negative result means either no localized extended 
emission present or emission that is clearly not associated with the object.

\item{SIMBAD associations:} These columns show the SIMBAD associations as described above.
In about a dozen cases, two probable associations are present, typically a planetary 
nebula coincident with our coordinates, plus an IRAS source that is almost certainly 
the same object but has not been noted as such in SIMBAD. In these cases
the primary identification is given as the planetary nebula or star, and the IRAS 
association is indicated in a footnote. Five objects were matched with separate
searches in the catalogs in the VizieR service which are not yet incorporated 
into SIMBAD; these are noted in footnotes. 

\end{description}

\section{Discussion}

While spectroscopic study would be necessary to determine the physical nature
of any given object, we may draw some tentative conclusions based on the known
identifications. It is striking that for the rings, disks, and two-lobed 
objects, nearly all of the specific SIMBAD identifications (50 of 57) are either 
PNe or PN candidates, and the remainder are identified as either stars or 
radio sources, which does not preclude these objects from being PNe as well. 
The objects in the catalog have been selected solely on the basis of their
24 {\micron} morphology, and without prior knowledge or expectation of what
any of these objects are, so it is tempting to conclude from the near-universality
of the PN identifications that the ring, disk, and two-lobed lists collectively
form a catalog of PNe in the MIPSGAL survey region, specifically PNe that are
both observed and resolved at the 6{\arcsec} 24 {\micron} MIPS resolution. 

This catalog could therefore contain up to 300 previously undiscovered PNe,
helping to alleviate the known discrepancy between the number of expected Galactic
PNe and the number that have been identified (e.g. Parker et al. 2003;
Phillips \& Ramos-Larios 2008), which can, at least in part, be attributed
to extinction effects that are largely absent at 24 \micron.
It is also possible, however, that the unidentified objects are located 
preferentially deeper in the Galactic disk, and some portion of them may be 
massive evolved stars or other object types observed from a great distance rather 
than garden-variety PNe.

The completeness of this tentative PN catalog (rings, disks, and two-lobed objects)
is limited by the general exclusion 
of centrally peaked objects, whether the selection of ``irregular'' objects 
encompasses the actual range of morphological PN variation, and the vagaries of
a visual search of a large image set. PNe can be strongly peaked at 24 \micron;
see Su et al. (2004) for an example that most likely would have been excluded from our 
catalog for that reason. Of the 35 ``peaked'' disks in the list,
all 7 with identifications are PNe, so this suggests that there are many more such objects
that have been omitted from the catalog. Circularly symmetric, centrally peaked 
extended objects are perhaps 10-15 times more numerous in the MIPSGAL data than the rings and disks
selected for this catalog, but this larger ensemble certainly includes objects 
such as YSOs (see, e.g., Cyganowski et al. 2008 for some examples) and unresolved
sources at 24 \micron.

In contrast to the ubiquitous identification of the ring and disk objects as
PNe, Morris et al. (2006) present an outer Galaxy object discovered in the {\spi} Galactic 
First Look Survey morphologically similar at 24 {\micron} to the objects 
selected for this catalog (we would categorize it as an irregular 
ring). Spectroscopy showed that this object lacks a dust continuum; virtually
all the 24 {\micron} emission is attributed to [O IV]. These authors interpret
this object as a young supernova remnant. This object is slightly larger
in angular extent, about 1{\arcmin}, than the largest of our rings and disks, 
but not enough to argue that it is in a separate class of objects. In other gross 
properties (low surface brightness, lack of detection in IRAC bands and 
70 \micron) it is similar to many of our ring and disk objects. Spectroscopic
data for one of our irregular ring objects, MGE059.4354-00.4662, also shows a lack of 
a dust continuum \citep{billot09}, and whose 24 {\micron} signature
is also due to ionized oxygen.

Chu et al. (2009) compared MIPS 24 {\micron} observations of Galactic PNe with archival
H$\alpha$ and He II data. They suggest that the 24 {\micron} emission is a combination
of dust continuum emission and the [O IV] line at 25.9 {\micron}, where the relative
contributions depend on the ionization and dust density structure of the PN in question.
Additional \spi Infrared Spectrometer (IRS) \citep{irs} observations of a small sample of 
ring and disk sources will help
determine the emission mechanisms in the present catalog (N. Billot et al. in preparation,
N. Flagey et al. in preparation).

The objects with central sources form an ensemble distinct from the sourceless
rings and disks. While two are identified as candidate PNe, the majority of the 
identifications (apart from the generic IR sources) are that of the central stars, and a 
preponderance of these are emission-line stars. 

While the visibility of
the residual star in a PN at 24 {\micron} is generally not anticipated, Su et al. (2007)
report a bright central pointlike source at 24 {\micron} in the Helix nebula, 
which they attribute to an unresolved debris disk surrounding the white
dwarf. The two identified central-source PNe candidates in our catalog fall in the angular size 
range of the rings and disks, and so this could be a plausible interpretation for the 
smaller of the unidentified central-source objects, but the majority of the 
central-source objects have a significantly larger angular size than even the 
largest rings and disks (or identified PNe among them), and so this interpretation 
should not be assumed for the central-source objects in general without further evidence.
(A debris disk model is more likely for the peaked disks in our catalog.)

We therefore attempt no general interpretation of the central-source
objects except to speculate that they are all evolved objects. One interesting
characteristic is their Galactic distribution: the apparent clustering at 
30{\dg} from the Galactic center corresponds to the tangent points of the
molecular ring. If these objects are largely confined to the molecular ring then
this suggests that they are evolved stages of massive, short-lived stars.

\acknowledgments

This work is based on observations made with the {\em Spitzer} Space 
Telescope, which is operated by the Jet Propulsion Laboratory, California 
Institute of Technology under a contract with NASA. Support for this work 
was provided by NASA in part through an award issued by JPL/Caltech. This 
research made use of the SIMBAD database and the Vizier catalog access tool, 
operated 
by the Centre de Donnees Astronomique de Strasbourg. This research has also 
made use of NASA's Astrophysics Data System Bibliographic Services.


\appendix
\section{Two possible new spiral galaxies}

As mentioned in the main text, in addition to the sources with a strong degree
of circular or bipolar symmetry, we also noted a number of other unusually
shaped objects. Of particular note are the
two objects shown in Figure \ref{sp3997x}. Each shows remarkable spiral structure at
both 24 and 8 \micron; neither has a counterpart found in SIMBAD.

For MGE314.2378+00.9793, the extended ($\sim45\times80\arcsec$ at 24 \micron)
spiral emission is prominent at 24 and 8 \micron, and faintly visible at 5.8
\micron. The central region appears dominated by 3.6 \micron\  emission,
although there is a central point source present in all 5 bands. In contrast,
the central region of MGE351.2381-00.0145 is dominated by a blob of 24 \micron\
emission ($\sim13\arcsec$) that is completely absent at the shorter
wavelengths. The spiral structure, $\sim75\arcsec$ across at 24 \micron, is
detected at 5.8 as well as 8  and 24 \micron. There also
appears to be 70 \micron\ emission associated with MGE351.2381-00.0145.

One possibility is that these are two spiral galaxies previously hidden due
to their location behind the Galactic plane. This seems particularly plausible
for MGE351.2381-00.0145, located at (l,b) = (351.24, -0.01);
MGE314.2378+00.9793 is at a slightly higher latitude, (l,b)= 314.24, +0.99),
but still well within the plane. Both are in rich star fields, as evidenced by
the numerous 3.6 \micron\ sources in the figure. MGE314.2378+00.9793 coincides
with an 843 MHz radio continuum source from the Sydney University Molongo Sky
Survey (Bock et al. 1999), and MGE351.2381-00.0145  has a radio continuum
source at 1.4 GHz from the NRAO VLA Sky Survey (NVSS, Condon et al. 1998).
The location of MGE314.2378+00.9793 is also coincident with the Norma Wall of
galaxies, about a degree from the two galaxies recently found with IRAC by the
GLIMPSE team (Jarrett et al. 2007). (Those two galaxies, MGE316.8732-00.5991
and 317.0392-00.4974, are actually two of the three objects in the
``miscellaneous'' group with SIMBAD associations.) Follow-up observations to
determine their radial velocities would be necessary to test this scenario
for the two spiral sources.

A second possibility for MGE314.2378+00.9793 is that there are two rings
superimposed, with one rotated about 30{\dg}. One trouble with this,
though, is that while there is a point source reasonably well-centered in the
overall structure, there do not seem to be point sources near the center of
either of the potential separate ring structures.

An alternative explanation for MGE351.2381-00.0145 is some kind of rotating
wind mechanism, creating a pinwheel nebula comparable to that around
Wolf-Rayet stars such as  WR 104 (Tuthill et al. 1999) or  WR 98a (Monnier
et al. 1999). The lack of a point source at the center does not really support
that, though, as well as the fact that the angular size of the MGE351.2381-00.0145
nebula, several hundred times the $\sim$0{\farcs}2 of WR 104, makes the WR interpretation unlikely. 
On the other hand, we know of no spiral galaxies
with such strong central emission at 24 \micron\ yet not at 8 \micron.

\begin{figure}
\epsscale{0.75}
\plotone{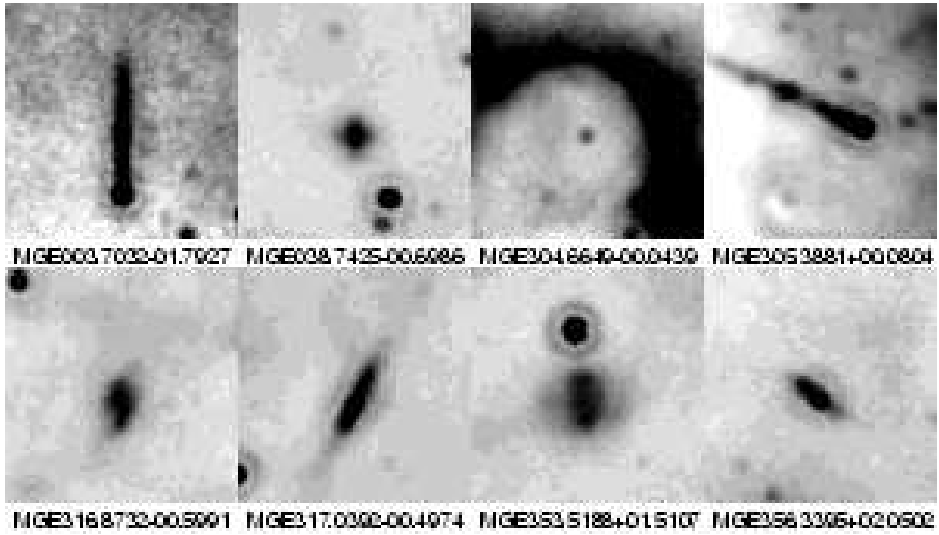}
\caption{Miscellaneous objects. Each panel is about 1{\farcm}5 across.
MGE316.8732-00.5991 and MGE317.0392-00.4974 are galaxies identified in the
GLIMPSE survey (Jarrett et al. 2007). 
}
\label{odd}
\end{figure}

\begin{figure}
\plotone{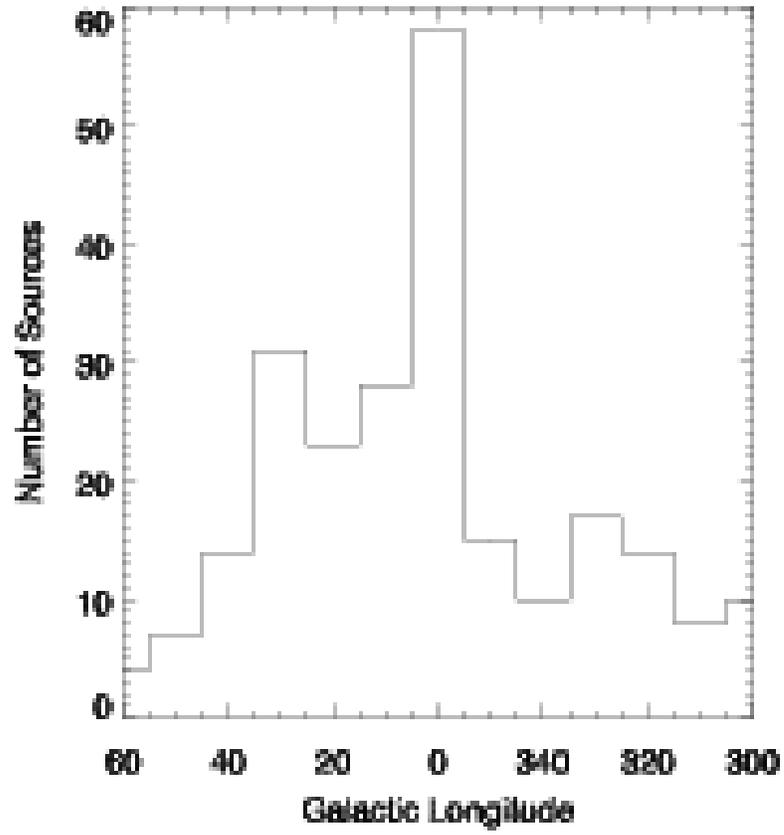}
\caption{Histogram of Galactic longitudes of objects within 1{\dg} of the plane.}
\label{galhist}
\end{figure}

\begin{figure}
\plotone{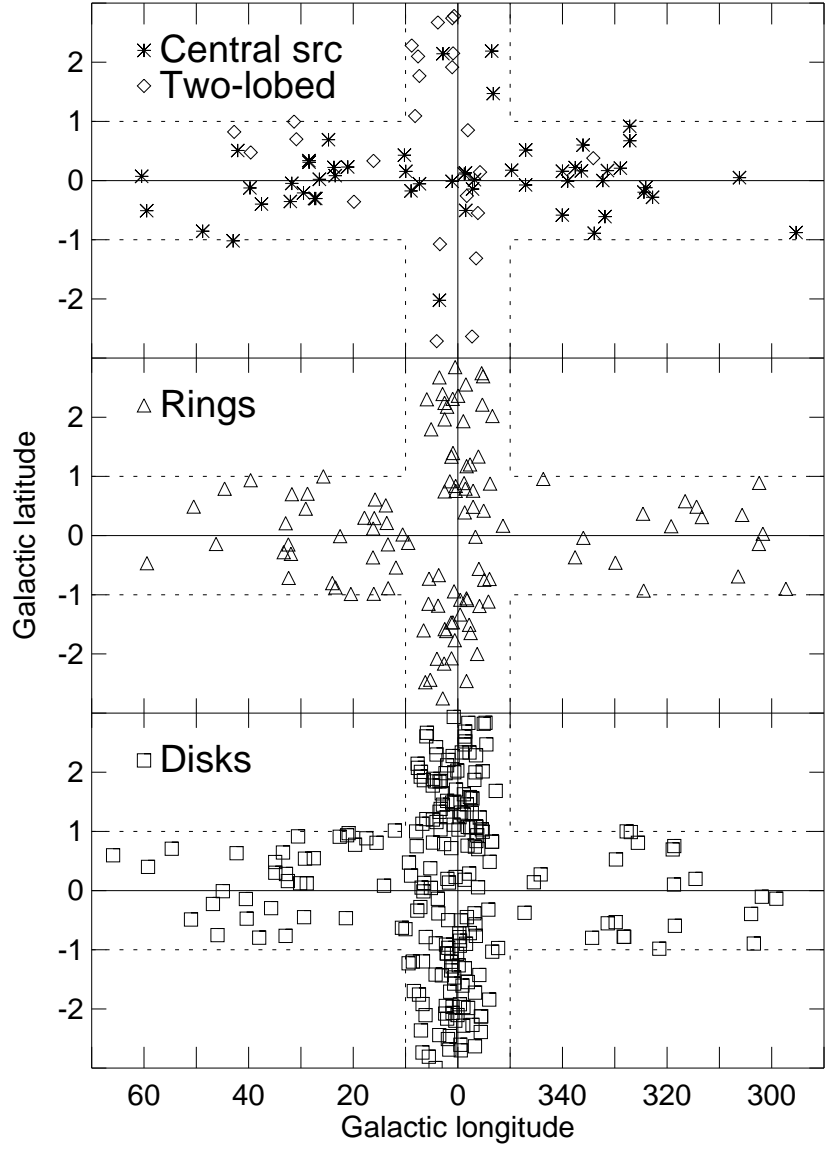}
\caption{Locations of the catalog objects in Galactic coordinates. The dashed lines
show the approximate limits of the MIPSGAL survey.}
\label{dist}
\end{figure}

\begin{figure}
\plotone{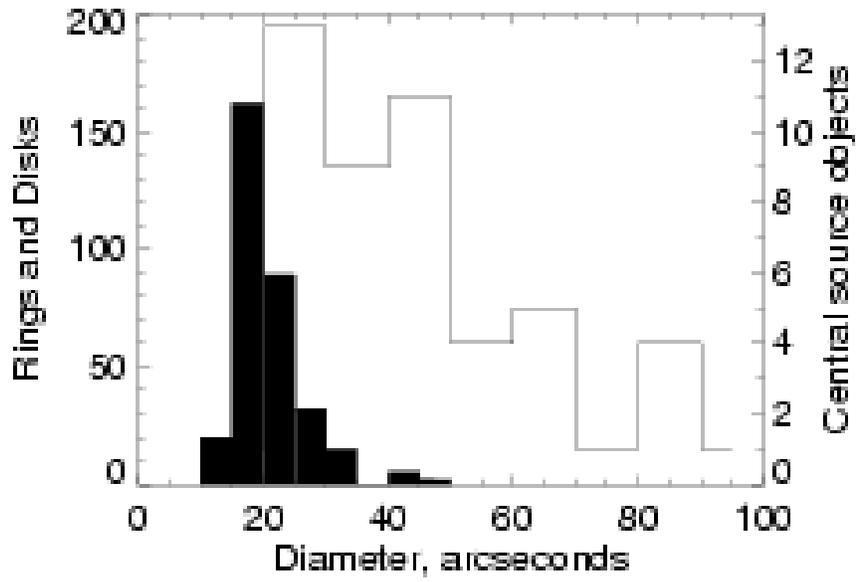}
\caption{Distribution of object diameters for central-source objects (open histogram) 
and rings and disks combined (filled histogram).}
\label{size}
\end{figure}

\begin{figure}
\epsscale{0.75}
\plotone{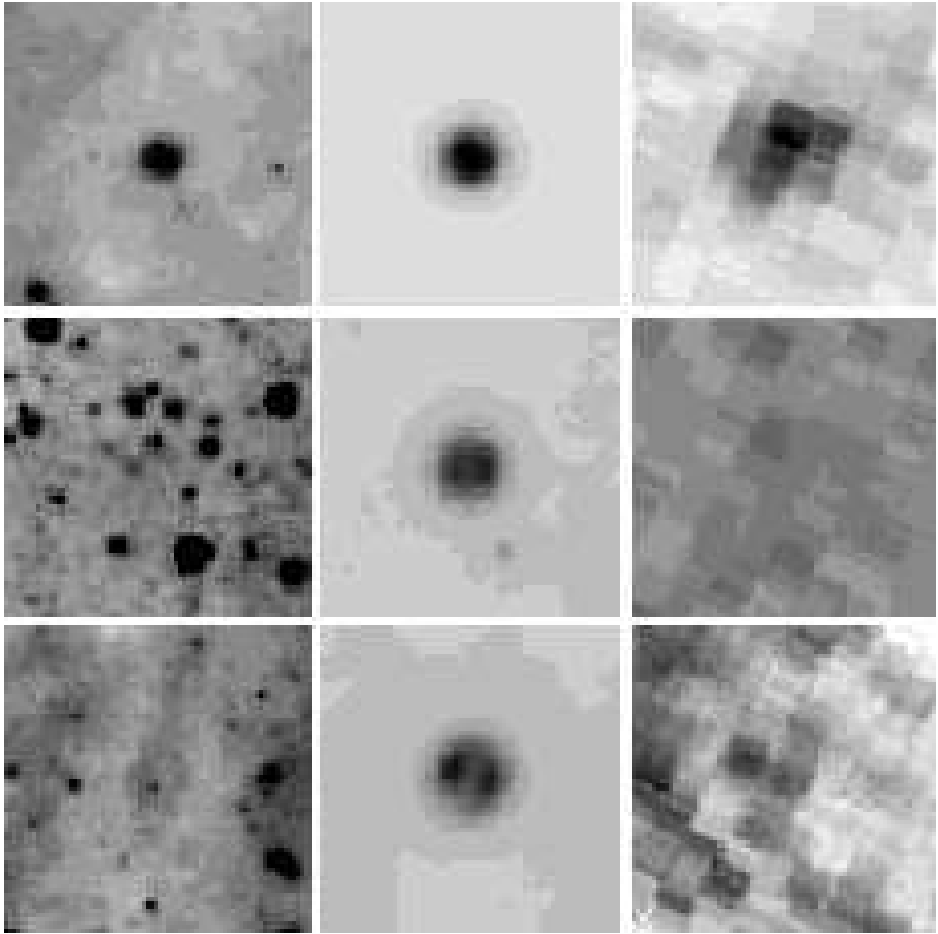}
\caption{Left to right: 8, 24, and 70 {\micron} images. Top: Definite 
detections at 8 and 70 {\micron} for flat disk MGE030.1503+00.1237. 
Center: Non-detections, bilaterally symmetric disk MGE002.2728-00.9131. 
Bottom: Ambiguous detections, bilaterally symmetric disk MGE007.7506-00.3392. 
Each panel is approximately 1{\farcm}3 across.}
\label{pan9}
\end{figure}

\begin{figure}
\plotone{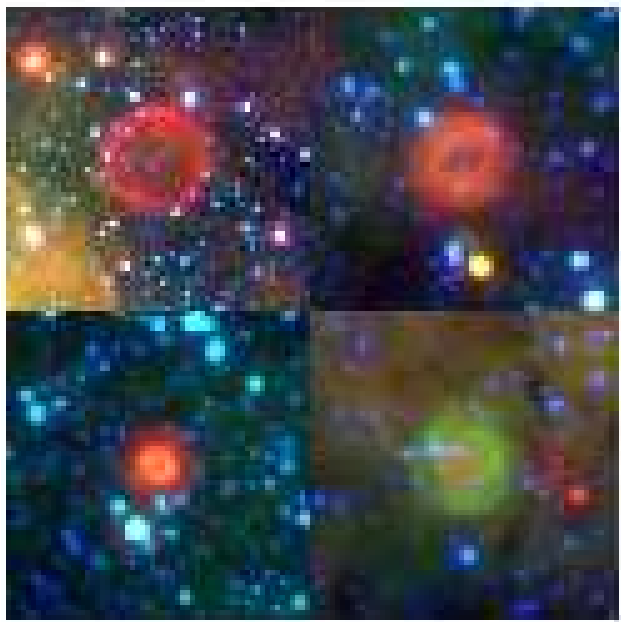}
\caption{Three-color images of four objects. Red: 24 \micron, green: 8 \micron, blue: 3.6 \micron.
Upper left: MGE015.8261+00.6109, no 8 \micron\ point source or
extended emission. Upper right: MGE029.0781+00.4547, 8 \micron\ emission
co-spatial with the 24 \micron\ emission. Lower left: MGE009.3521+00.4736,
a thin ring of 8 \micron\ emission interior to the 24 \micron\ emission.
Lower right: MGE337.5950-00.3664, the only instance where the 8 \micron\
emission surrounds the 24 \micron\ structure. The image size for MGE015.8261+00.6109 is 
4{\farcm}2${\times}$4{\farcm}2; the other
three are
2{\farcm}1${\times}$2{\farcm}1.}
\label{fig.3c}
\end{figure}

\begin{figure}
\plottwo{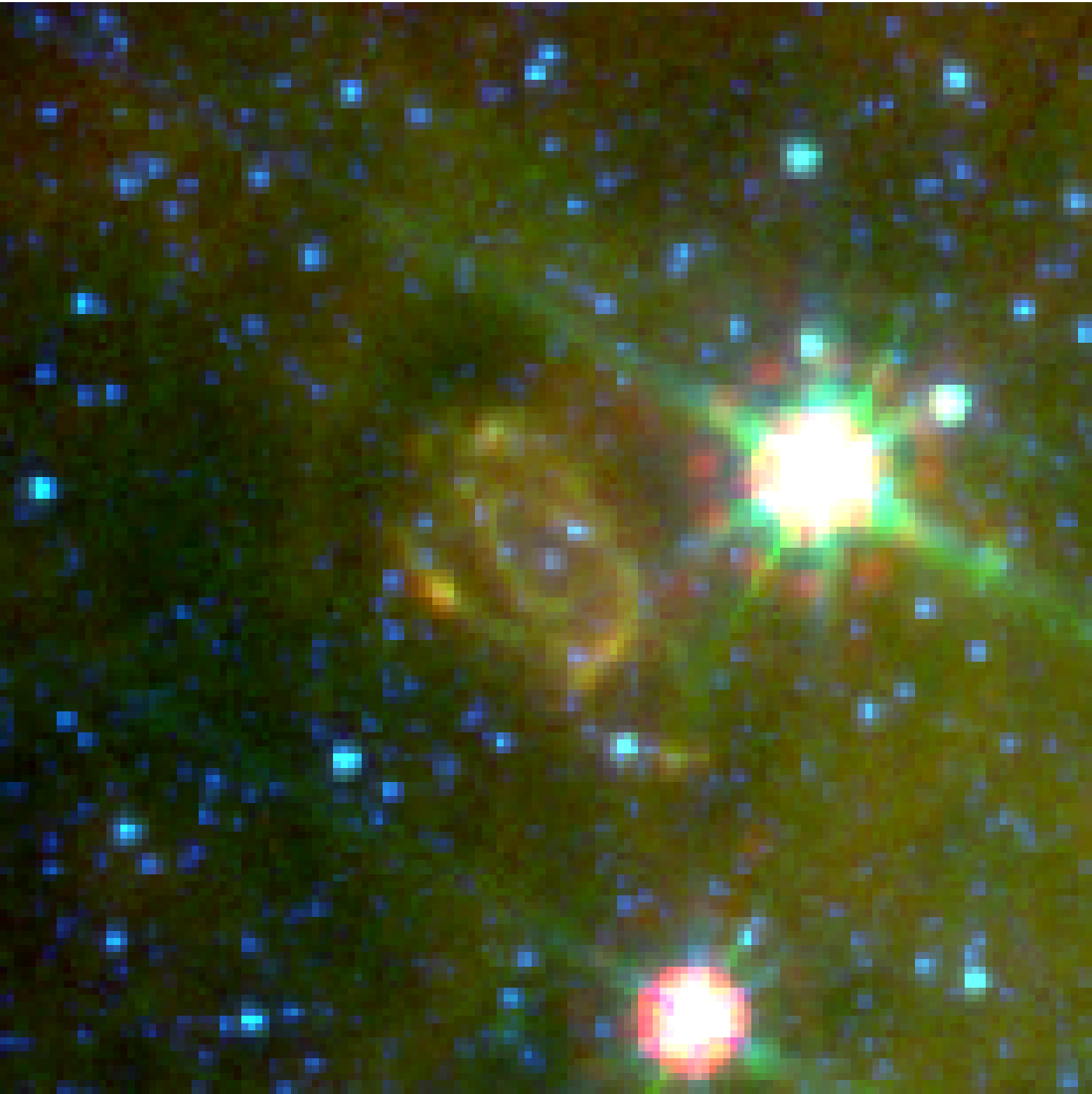}{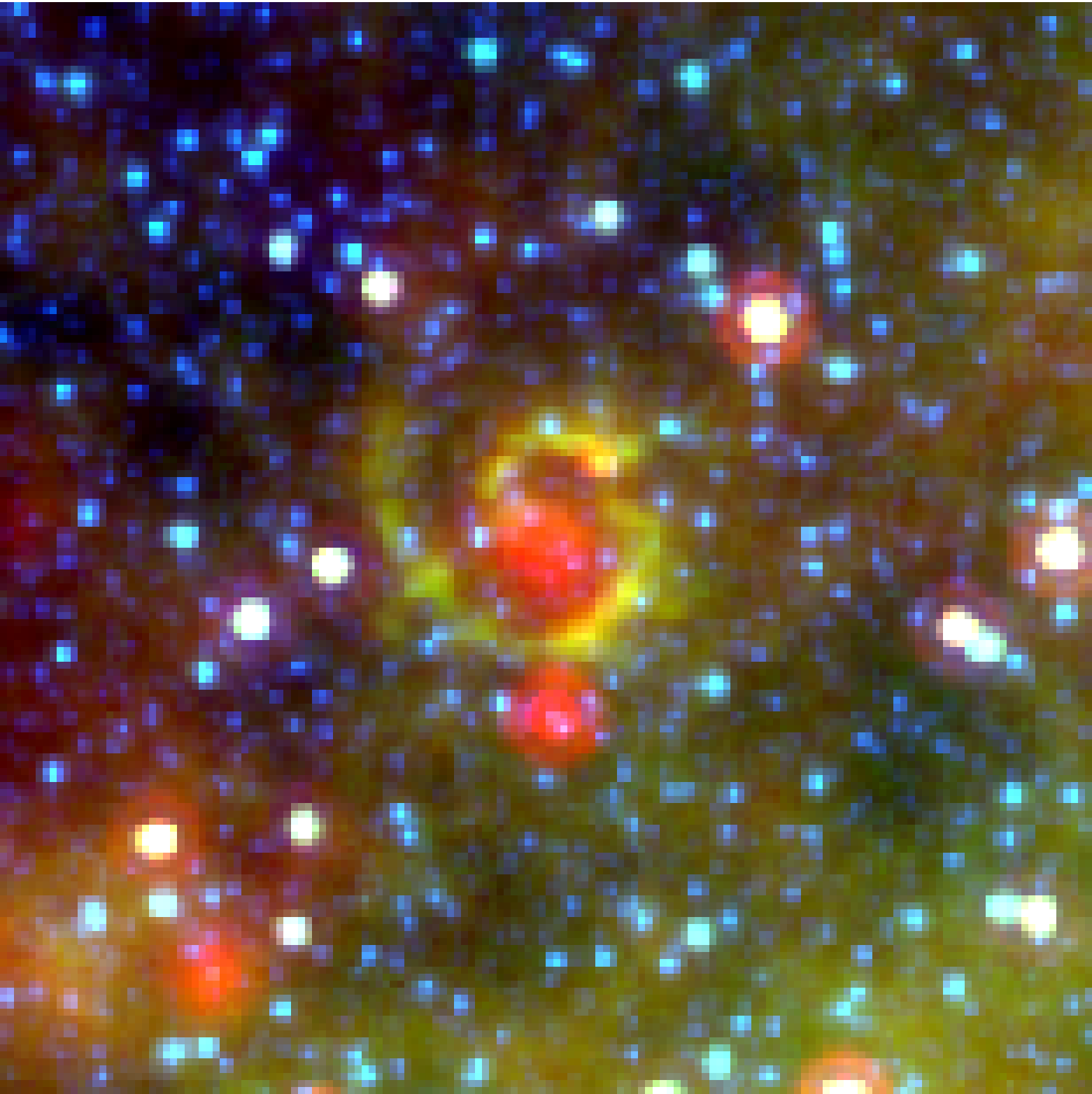}
\caption{Left: Three-color image of MGE314.2378+00.9793. Red: 24 \micron,
green: 8 \micron, blue: 3.6 \micron.  There is 8 and 24 {\micron} emission
from the central source but it is dominated by the 3.6 {\micron} emission
with this stretch. 
The bright star to the right is IRAS 14195-5938.
Right: Three-color image of MGE351.2381-00.0145.}
\label{sp3997x}
\end{figure}

Another ``miscellaneous'' object, MGE356.3395+02.0502, shown in Figure \ref{odd},
is a central point source with a bright bar
of emission plus a fainter halo above and below the bar. It, too, is an NVSS
point source, and could be a smaller or more distant spiral seen edge-on,
rather than face-on as with the first two objects. Unfortunately, at
(l,b)$\sim$356.34, 2.05, it was outside the GLIMPSE coverage. Again, additional
observations are needed to determine what this source actually is.

MGE305.3881+00.0804 has a morphological similarity to photoablating proplyd candidates
observed at optical wavelengths in Orion (O'Dell et al. 1993) and the Carina Nebula (Smith et al. 2003).
The Wolf-Rayet star WR 48a is about 2{\arcmin} distant and is approximately aligned with the head-tail
orientation of this object. At a distance of 4 kpc, the $\sim$1{\arcmin} tail is $\sim$1 pc in length, 
$\sim$100 times that of the Orion proplyds and $\sim$10 times those in Carina, which suggests that this
could be a much more massive protoplanetary disk than those found in Orion or Carina.

MGE003.7032-01.7927 resembles a proplyd, but there is no apparent source driving the flow. A jet could
be a possibility, but this is nothing like a YSO protostellar jet, and there is no signature of a counter-jet.

\clearpage

\newpage

\newpage

\newpage

\newpage

\newpage

\newpage

\newpage

\newpage

\newpage

\newpage

\clearpage

\begin{deluxetable}{lrrcccccccll}
\tabletypesize\tiny
\tablewidth{0pt}
\tablecaption{MIPSGAL disk and ring catalog\label{tab.catalog2}}
\tablecolumns{12}
\tablehead{
\colhead{Name} & 
    \multicolumn{2}{c}{J2000 Coordinates} & 
    \colhead{Diam.} & 
    \colhead{F$_{24}$} & 
    \multicolumn{3}{c}{Detection Flags} &  
    \colhead{} & 
    \multicolumn{3}{c}{SIMBAD Associations}\\
\cline{2-3} \cline{6-8} \cline{10-12} \\
\colhead{} & 
    \colhead{$\alpha(^{\mbox{h}} \hspace{1mm} ^{\mbox{m}} \hspace{1mm} ^{\mbox{s}})$} & 
    \colhead{$\delta(\arcdeg \hspace{1mm} \arcmin \hspace{1mm} \arcsec)$} & 
    \colhead{(\arcsec)} & 
    \colhead{(Jy)} & 
    \colhead{8\micron} &  
    \colhead{8\micron} &
    \colhead{70\micron} & 
    \colhead{} & 
    \colhead{Dist.} & 
    \colhead{Name} & 
    \colhead{Object} \\
\colhead{} & 
    \colhead{} & 
    \colhead{} & 
    \colhead{} &
    \colhead{} & 
    \colhead{Point} & 
    \colhead{Ext.} & 
    \colhead{Ext.} & 
    \colhead{} & 
    \colhead{(\arcsec)} &
    \colhead{} & 
    \colhead{Type} \\
\colhead{(1)} & 
    \colhead{(2)} & 
    \colhead{(3)} & 
    \colhead{(4)} &
    \colhead{(5)} & 
    \colhead{(6)} & 
    \colhead{(7)} & 
    \colhead{(8)} & 
    \colhead{} & 
    \colhead{(9)} &
    \colhead{(10)} & 
    \colhead{(11)} 
}
\startdata
\multicolumn{12}{c}{1a: Objects with central sources---regular}\\\hline
MGE002.8493+02.1430 & 17 44 05.3 & --25 23 16 & 24 & 0.04(15) & -- & -- & ? &&  &  & \\
MGE003.5216-02.0237 & 18 01 36.1 & --26 55 40 & 21 & 0.04(24) & -- & -- & N &&  &  & \\
MGE007.3429-00.0549 & 18 02 22.3 & --22 38 00 & 32 & 1.1(80) & Y & Y & -- &&  &  & \\
MGE008.9460-00.1750 & 18 06 13.1 & --21 17 45 & 15 & 0.13(35) & Y & N & ? && 1.5 & HD 313771 & *\\
MGE009.9541+00.1556 & 18 07 05.2 & --20 15 16 & 28 & 0.35(18) & Y & ? & ? &&  &  & \\
MGE010.2114+00.4289 & 18 06 36.2 & --19 53 48 & 40 & 18.($<1$) & Y & Y\tablenotemark{k} & Y && 4.3 & IRAS 18036-1954 & PN?\\
MGE021.0510+00.2292 & 18 28 41.5 & --10 27 06 & 45 & 0.29(99) & Y & N & N &&  &  & \\
MGE023.4499+00.0820 & 18 33 43.4 & --08 23 35 & 25 & 1.7(34) & Y & N & ? &&  &  & \\
MGE023.6857+00.2226 & 18 33 39.5 & --08 07 08 & 44 & 3.7(29) & Y & N & ? &&  &  & \\
MGE024.7290+00.6910 & 18 33 55.3 & --06 58 38 & 50 & 11.(4) & Y & Y & Y && 1.1 & V* V481 Sct & LBV\\
MGE027.1745-00.3041 & 18 41 59.6 & --05 15 41 & 32 & 2.6(3) & Y & ? & Y &&  &  & \\
MGE027.3839-00.3031 & 18 42 22.5 & --05 04 29 & 38 & 11.(2) & Y & N & Y && 12.2 & IRAS 18397-0507 & IR\\
MGE029.5086-00.2090 & 18 45 55.9 & --03 08 30 & 34 & 7.0(5) & Y & Y & Y &&  &  & \\
MGE031.7516-00.0479 & 18 49 27.3 & --01 04 20 & 240 & 8.0(99) & Y & N & ? &&  & \tablenotemark{j} & \\
MGE037.5416-00.3975 & 19 01 16.6 & +03 55 11 & 42 & 1.1(19) & Y & ? & ? && 45.5 & IRAS 18588+0350\tablenotemark{e} & IR\\
MGE039.7412-00.1246 & 19 04 20.9 & +06 00 00 & 46 & 0.34(39) & Y & N & N &&  &  & \\
MGE042.0787+00.5084 & 19 06 24.5 & +08 22 02 & 21 & 3.4($<1$) & Y & N & Y && 1.6 & IRAS 19040+0817 & IR\\
MGE042.9679-01.0183 & 19 13 32.7 & +08 27 04 & 38 & 0.12(59) & Y & N & N &&  &  & \\
MGE048.7815-00.8565 & 19 24 03.3 & +13 39 50 & 30 & 0.14(16) & Y & N & ? &&  &  & \\
MGE059.4884-00.5093 & 19 44 42.9 & +23 11 34 & 48 & 0.18(20) & Y & N & N &&  &  & \\
MGE060.4530+00.0712 & 19 44 37.5 & +24 19 06 & 55 & 0.84(11) & Y & N & ? && 7.3 & HBHA 2203-01\tablenotemark{b} & Em*\\
MGE295.3343-00.8786 & 11 44 17.8 & --62 45 21 & 40 & 1.6(10) & Y & ? & Y && 3.5 & IRAS 11419-6228 & IR\\
MGE306.1565+00.0494 & 13 19 33.9 & --62 38 46 & 26 & 0.99(16) & Y & N & N && 1.0 & CD-61 3738 & *\\
MGE322.7707-00.2829 & 15 25 59.7 & --57 04 40 & 22 & 0.04(20) & Y & Y & Y &&  &  & \\
MGE324.0739-00.1200 & 15 33 08.2 & --56 12 19 & 45 & 1.2(17) & Y & N & ? &&  & \tablenotemark{i} & \\
MGE324.4138-00.1964 & 15 35 26.5 & --56 04 13 & 64 & 0.54(89) & Y & N & Y &&  &  & \\
MGE327.1444+00.9168 & 15 45 59.1 & --53 32 33 & 53 & 1.4(19) & Y & N & -- && 10.2 & IRAS 15421-5323 & IR\\
MGE328.9301+00.2092 & 15 58 13.9 & --52 57 51 & 55 & 2.5(99) & Y & N & ? &&  &  & \\
MGE331.3971+00.1685 & 16 10 26.5 & --51 21 25 & 25 & 1.0(19) & Y & N & ? &&  &  & \\
MGE332.2843-00.0002 & 16 15 17.7 & --50 52 18 & 47 & 106.(--)\tablenotemark{l} & Y & N & Y && 3.4 & IRAS 16115-5044 & pA*\\
MGE333.9204-00.8912 & 16 26 34.2 & --50 21 01 & 25 & 9.2(2) & Y & N & Y && 22.3 & IRAS 16228-5014\tablenotemark{g} & PN?\\
MGE336.0610+00.6001 & 16 29 03.8 & --47 46 25 & 33 & 6.2(2) & Y & Y & Y && 16.2 & IRAS 16254-4739\tablenotemark{f} & IR\\
MGE337.5544+00.2198 & 16 36 42.8 & --46 56 21 & 28 & 7.5(6) & Y & ? & Y &&  &  & \\
MGE338.9975-00.0082 & 16 43 16.2 & --46 00 42 & 60 & 25.(18) & Y & Y & ? && 3.7 & Wray 16-232 & Em*\\
MGE339.9975+00.1564 & 16 46 17.3 & --45 08 48 & 24 & 2.0(24) & Y & N & ? &&  &  & \\
MGE340.0297-00.5840 & 16 49 37.7 & --45 36 00 & 85 & 8.9(89) & Y & N & Y &&  &  & \\
MGE347.0305+00.5154 & 17 08 29.3 & --39 25 15 & 45 & 0.77(32) & Y & N & Y && 4.4 & IRAS 17050-3921 & IR\\
MGE347.0516-00.0746 & 17 11 00.9 & --39 45 17 & 38 & 5.0(21) & Y & N & ? &&  &  & \\
MGE353.2347+01.4703 & 17 22 36.8 & --33 49 09 & 24 & 0.13(16) & -- & -- & N &&  &  & \\
MGE353.5303+02.1881 & 17 20 35.1 & --33 10 06 & 225 & 5.1(27) & -- & -- & Y && 45.0 & Wray 15-680\tablenotemark{a} & Em*\\
MGE356.8235+00.0139 & 17 37 47.5 & --31 37 34 & 40 & 1.9(15) & Y & Y & -- &&  &  & \\
MGE357.1829-00.1434 & 17 39 19.0 & --31 24 22 & 66 & 2.9(18) & Y & N & -- &&  &  & \\
MGE358.5405+00.1299 & 17 41 35.4 & --30 06 38 & 80 & 69.(15) & Y & Y & -- && 1.4 & Wray 17-96 & LBV\\
MGE358.7700+00.1088 & 17 42 14.0 & --29 55 36 & 25 & 1.2(10) & Y & Y & -- &&  &  & \\
\cutinhead{1b: Objects with central sources---irregular}
MGE001.0860-00.0132 & 17 48 14.0 & --28 00 52 & 90 & 61.(28) & Y & ? & -- && 1.1 & MWC 272 & Em*\\
MGE026.4700+00.0209 & 18 39 32.2 & --05 44 20 & 80 & 86.(2) & Y & Y & Y && 1.2 & 2MASS${\dots}$\tablenotemark{d} & LBV\\
MGE028.4451+00.3094 & 18 42 08.2 & --03 51 03 & 80 & 20.(22) & Y & ? & Y &&  &  & \\
MGE028.4812+00.3368 & 18 42 06.3 & --03 48 23 & 240 & 9.0(99) & Y & ? & ? &&  &  & \\
MGE032.0204-00.3538 & 18 51 02.1 & --00 58 21 & 66 & 3.3(55) & ? & N & ? &&  &  & \\
MGE327.1201+00.6731 & 15 46 52.1 & --53 44 58 & 92 & 0.50(29) & N & N & ? &&  &  & \\
MGE331.8690-00.6096 & 16 16 04.6 & --51 35 56 & 30 & 42.(--)\tablenotemark{m} & Y & ? & Y && 3.3 & IRAS 16122-5128 & IR\\
MGE336.3749+00.1647 & 16 32 13.9 & --47 50 39 & 70 & 8.0(30) & ? & N & ? &&  &  & \\
MGE349.7294+00.1747 & 17 17 59.3 & --37 26 09 & 68 & 44.(6) & Y & Y & Y && 29.8 & SNR G349.7+00.2\tablenotemark{b} & SNR\\
MGE358.5016-00.5048 & 17 43 59.8 & --30 28 39 & 140 & 14.(--) & Y & ? & -- && 2.3 & [RHI84] 10-469\tablenotemark{b} & *\\
\cutinhead{2a: Rings}
MGE000.4176+00.7559 & 17 43 40.7 & --28 11 06 & 15 & 0.13(7) & ? & N & -- &&  &  & \\
MGE000.5194+02.8468 & 17 35 55.6 & --26 59 18 & 22 & 0.17(5) & -- & -- & ? && 1.5 & PN G000.5+02.8 & PN\\
MGE000.7235-00.9420 & 17 51 00.6 & --28 48 10 & 15 & 0.01(88) & ? & N & N &&  &  & \\
MGE000.9363+01.3962 & 17 42 26.9 & --27 24 26 & 16 & 0.12(15) & N & N & N &&  &  & \\
MGE001.2124-02.0748 & 17 56 36.2 & --28 57 18 & 18 & 0.04(13) & N & N & ? && 3.7 & PN G001.2-02.0\tablenotemark{b} & PN?\\
MGE002.2128-01.6131 & 17 57 03.9 & --27 51 30 & 18 & 0.26(11) & ? & N & ? &&  &  & \\
MGE002.5021-01.5808 & 17 57 35.6 & --27 35 31 & 18 & 0.08(10) & ? & N & N &&  &  & \\
MGE002.5220+01.9643 & 17 44 00.3 & --25 45 36 & 18 & 0.15(3) & -- & -- & Y && 1.2 & PHR J1744-2545 & PN?\\
MGE003.4454-00.6454 & 17 56 05.2 & --26 18 24 & 63 & 1.6($<1$) & N & N & ? &&  &  & \\
MGE003.6547-00.6687 & 17 56 38.7 & --26 08 15 & 18 & 0.12(6) & N & N & N &&  &  & \\
MGE003.9942-02.0816 & 18 02 52.2 & --26 32 43 & 18 & 0.07(17) & -- & -- & ? &&  &  & \\
MGE005.5058-00.7281 & 18 00 57.6 & --24 33 47 & 30 & 3.3(2) & Y & ? & Y && 3.2 & IRAS 17578-2433 & IR\\
MGE005.6102-01.1516 & 18 02 48.4 & --24 40 54 & 20 & 2.3(4) & N & Y & Y && 4.3 & IRAS 17597-2441 & IR\\
MGE006.2233-02.4777 & 18 09 13.3 & --24 47 36 & 28 & 0.05(36) & -- & -- & N &&  &  & \\
MGE010.5569+00.0188 & 18 08 50.5 & --19 47 38 & 25 & 0.95(8) & N & N & ? && 30.0 & IRAS 18058-1948 & IR\\
MGE013.3662-00.1483 & 18 15 09.7 & --17 24 40 & 21 & 0.44(25) & N & N & ? &&  &  & \\
MGE013.7676+00.5111 & 18 13 32.5 & --16 44 36 & 18 & 0.32(14) & N & N & ? &&  &  & \\
MGE015.8263+00.6108 & 18 17 15.4 & --14 53 10 & 45 & 0.62(22) & N & N & ? &&  &  & \\
MGE016.1871+00.1202 & 18 19 45.1 & --14 48 02 & 16 & 0.05(32) & N & N & N &&  &  & \\
MGE023.3894-00.8753 & 18 37 02.9 & --08 53 14 & 30 & 0.36(18) & N & N & ? &&  &  & \\
MGE024.0055-00.8031 & 18 37 56.0 & --08 18 24 & 21 & 0.08(35) & N & N & ? &&  &  & \\
MGE031.9075-00.3087 & 18 50 40.1 & --01 03 09 & 25 & 1.4(7) & N & N & ? && 4.9 & PK 031-00 1 & PN\\
MGE039.6131+00.9358 & 19 00 19.2 & +06 22 18 & 16 & 0.06(7) & N & N & N &&  &  & \\
MGE301.6857+00.0304 & 12 40 31.4 & --62 48 54 & 17 & 0.05(7) & N & N & ? &&  &  & \\
MGE302.3971+00.8908 & 12 46 53.1 & --61 58 34 & 13 & 0.01(66) & N & N & -- &&  &  & \\
MGE306.4142-00.6894 & 13 22 34.0 & --63 21 02 & 32 & 2.7(2) & N & Y & Y && 0.1 & Wray 16-127 & PN\\
MGE316.5509+00.5814 & 14 41 27.8 & --59 20 58 & 24 & 0.98(2) & N & N & ? &&  &  & \\
MGE337.5950-00.3666 & 16 39 26.1 & --47 18 04 & 24 & 0.20(41) & Y & Y & ? &&  &  & \\
MGE354.1474-01.1141 & 17 35 27.6 & --34 29 19 & 40 & 0.49(14) & N & N & -- &&  &  & \\
MGE355.1739+02.6929 & 17 23 01.6 & --31 31 47 & 24 & 0.15(7) & -- & -- & N &&  &  & \\
MGE356.0690+01.3365 & 17 30 39.1 & --31 32 32 & 18 & 0.04(20) & N & N & N &&  &  & \\
MGE357.5633-01.6468 & 17 46 15.8 & --31 52 25 & 20 & 0.06(19) & N & N & N &&  &  & \\
MGE357.6735+01.2030 & 17 35 15.0 & --30 16 11 & 21 & 0.15(7) & ? & N & -- &&  &  & \\
MGE358.2249-01.0885 & 17 45 38.4 & --31 01 06 & 15 & 0.04(26) & N & N & N &&  &  & \\
MGE358.3271+01.1733 & 17 36 59.4 & --29 44 07 & 26 & 0.19(6) & N & N & N &&  &  & \\
MGE358.3478-01.0603 & 17 45 49.6 & --30 53 55 & 22 & 0.03(38) & N & N & N &&  &  & \\
MGE358.4694+02.5544 & 17 32 01.1 & --28 52 05 & 24 & 0.05(14) & -- & -- & N &&  &  & \\
MGE358.6905+00.3921 & 17 40 55.8 & --29 50 40 & 15 & 0.08(22) & N & N & -- && 4.5 & [DVC98] 6 & Rad\\
\cutinhead{2b: Rings---irregular}
MGE000.4100+00.8386 & 17 43 20.5 & --28 08 53 & 18 & 0.21(12) & N & N & ? &&  &  & \\
MGE000.5910-01.7683 & 17 53 57.5 & --29 20 14 & 25 & 0.25(7) & ? & N & N && 4.7 & JaSt 96 & PN\\
MGE000.9516-01.4656 & 17 53 35.9 & --28 52 23 & 18 & 0.12(14) & N & N & ? && 4.2 & JaSt2 13 & PN\\
MGE000.9899+02.3115 & 17 39 05.1 & --26 52 39 & 21 & 0.27(4) & -- & -- & ? &&  &  & \\
MGE001.2343+01.3341 & 17 43 23.4 & --27 11 12 & 25 & 0.13(11) & ? & N & N && 5.3 & JaSt 45 & PN\\
MGE001.2920-01.4680 & 17 54 23.6 & --28 34 51 & 15 & 0.03(14) & N & N & N && 0.4 & JaSt2 15 & PN\\
MGE001.5280+00.9171 & 17 45 40.7 & --27 09 15 & 18 & 0.20(7) & N & N & ? &&  &  & \\
MGE002.0644+02.1789 & 17 42 07.5 & --26 02 12 & 30 & 0.05(33) & -- & -- & ? &&  &  & \\
MGE002.5276+00.7462 & 17 48 39.2 & --26 23 16 & 18 & 0.18(9) & ? & N & -- &&  &  & \\
MGE002.5424+02.2426 & 17 43 00.0 & --25 35 49 & 25 & 0.05(18) & -- & -- & ? &&  &  & \\
MGE002.6193-02.1625 & 18 00 08.1 & --27 46 48 & 23 & 0.02(51) & -- & -- & N &&  &  & \\
MGE002.9176+02.3920 & 17 43 18.5 & --25 11 57 & 25 & 0.01(42) & -- & -- & N &&  &  & \\
MGE003.5282+02.6745 & 17 43 39.3 & --24 31 54 & 50 & 0.69(3) & -- & -- & Y && 1.6 & PN G003.5+02.7\tablenotemark{b} & PN\\
MGE003.7724-01.1833 & 17 58 53.6 & --26 17 35 & 28 & 0.30(8) & N & N & ? &&  &  & \\
MGE005.2835-02.4370 & 18 07 03.3 & --25 35 43 & 25 & 0.46(4) & -- & -- & ? && 1.2 & PHR J1807-2535 & PN\\
MGE005.9347+02.3035 & 17 50 28.1 & --22 39 57 & 30 & 0.05(15) & -- & -- & N &&  &  & \\
MGE006.5449-01.6013 & 18 06 32.4 & --24 05 14 & 34 & 0.29(33) & -- & -- & N &&  &  & \\
MGE009.4947-00.1225 & 18 07 10.1 & --20 47 28 & 20 & 0.15(15) & N & N & N &&  &  & \\
MGE011.8322-00.5388 & 18 13 31.1 & --18 56 42 & 26 & 0.47(14) & Y & N & Y &&  &  & \\
MGE013.3211-00.8896 & 18 17 48.5 & --17 48 12 & 18 & 0.32(6) & N & Y & N &&  &  & \\
MGE015.9774+00.2955 & 18 18 42.2 & --14 54 09 & 18 & 0.18(51) & ? & N & ? &&  &  & \\
MGE016.0749-00.9833 & 18 23 34.1 & --15 25 07 & 25 & 0.13(42) & N & N & -- &&  &  & \\
MGE016.2280-00.3680 & 18 21 36.9 & --14 59 41 & 18 & 0.30(35) & ? & N & ? && 3.5 & GPSR5 16.228-0.369 & Rad\\
MGE017.8638+00.3028 & 18 22 20.6 & --13 14 08 & 20 & 0.15(23) & ? & N & ? &&  &  & \\
MGE020.4513-00.9867 & 18 31 57.1 & --11 32 46 & 32 & 0.35(14) & N & N & -- &&  &  & \\
MGE025.7137+01.0007 & 18 34 38.4 & --05 57 38 & 17 & 0.04(61) & N & N & ? &&  &  & \\
MGE028.7440+00.7076 & 18 41 16.0 & --03 24 11 & 23 & 0.25(10) & N & N & ? &&  &  & \\
MGE031.7290+00.6993 & 18 46 45.2 & --00 45 06 & 44 & 1.2(8) & ? & Y\tablenotemark{k} & Y && 12.7 & IRAS 18441-0048 & IR\\
MGE032.3506-00.7153 & 18 52 55.4 & --00 50 37 & 23 & 0.11(34) & N & N & ? &&  &  & \\
MGE032.4006-00.1507 & 18 51 00.3 & --00 32 30 & 33 & 0.69(7) & Y & N & ? &&  &  & \\
MGE046.2242-00.1416 & 19 16 32.6 & +11 44 30 & 25 & 0.11(23) & N & N & N &&  &  & \\
MGE050.5016+00.4896 & 19 22 30.4 & +15 48 56 & 28 & 0.33(5) & ? & N & ? &&  &  & \\
MGE059.4354-00.4662 & 19 44 26.3 & +23 10 06 & 25 & 0.07(21) & N & N & N &&  &  & \\
MGE297.2836-00.8995 & 12 00 58.8 & --63 13 00 & 23 & 0.06(18) & ? & N & N && 12.3 & HD 104375 & *\\
MGE302.4617-00.1412 & 12 47 17.6 & --63 00 33 & 20 & 0.15(23) & N & N & ? &&  &  & \\
MGE305.6516+00.3494 & 13 14 56.9 & --62 23 53 & 19 & 0.59(33) & Y & ? & N &&  &  & \\
MGE313.3544+00.3123 & 14 18 27.6 & --60 47 11 & 24 & 0.81(3) & N & Y & ? && 0.2 & PN G313.3+00.3\tablenotemark{b} & PN\\
MGE314.3603+00.4880 & 14 25 39.2 & --60 16 35 & 20 & 0.08(29) & ? & Y & ? &&  &  & \\
MGE319.2193+00.1581 & 15 01 27.5 & --58 32 26 & 26 & 1.3(4) & N & N & ? &&  &  & \\
MGE324.4811-00.9290 & 15 38 56.9 & --56 37 21 & 18 & 0.03(99) & N & N & ? &&  &  & \\
MGE324.6051+00.3692 & 15 34 12.4 & --55 29 56 & 25 & 0.21(7) & ? & Y & Y &&  &  & \\
MGE329.8780-00.4579 & 16 05 52.6 & --52 50 35 & 40 & 0.09(99) & N & N & N &&  &  & \\
MGE336.0114-00.0381 & 16 31 37.9 & --48 14 54 & 28 & 4.3(8) & Y & N & Y && 28.4 & IRAS 16278-4808\tablenotemark{h} & IR\\
MGE343.6641+00.9584 & 16 55 46.9 & --41 48 42 & 21 & 0.13(22) & ? & N & N &&  &  & \\
MGE351.3662+00.1671 & 17 22 41.4 & --36 05 50 & 18 & 0.11(13) & ? & N & N &&  &  & \\
MGE353.3934+02.0206 & 17 20 52.2 & --33 22 34 & 18 & 0.17(6) & -- & -- & N &&  &  & \\
MGE353.8415+00.8789 & 17 26 36.4 & --33 39 00 & 15 & 0.13(14) & N & N & ? &&  &  & \\
MGE353.9398-00.7336 & 17 33 21.8 & --34 27 24 & 24 & 0.47(4) & N & N & -- &&  &  & \\
MGE354.9589-00.7468 & 17 36 04.8 & --33 36 26 & 15 & 0.15(11) & N & N & N &&  &  & \\
MGE355.0606+00.4209 & 17 31 39.4 & --32 53 15 & 15 & 0.27(4) & ? & N & ? &&  &  & \\
MGE355.3208+02.2134 & 17 25 16.6 & --31 40 40 & 23 & 0.05(23) & -- & -- & -- &&  &  & \\
MGE355.4464+02.7440 & 17 23 32.9 & --31 16 34 & 20 & 0.01(78) & -- & -- & N &&  &  & \\
MGE355.8710-01.1891 & 17 40 12.3 & --33 04 24 & 24 & 0.08(18) & N & N & N &&  &  & \\
MGE355.9808-00.5628 & 17 37 57.6 & --32 38 48 & 24 & 0.09(25) & ? & N & -- &&  &  & \\
MGE356.3082-02.0001 & 17 44 35.8 & --33 07 43 & 25 & 0.09(10) & -- & -- & N &&  &  & \\
MGE356.6350-00.0191 & 17 37 26.8 & --31 48 10 & 18 & 0.06(13) & Y & Y & -- &&  &  & \\
MGE357.0783+00.4869 & 17 36 33.7 & --31 09 26 & 15 & 0.11(23) & N & N & -- &&  &  & \\
MGE357.0801+00.7581 & 17 35 30.0 & --31 00 34 & 15 & 0.02(37) & ? & N & N &&  &  & \\
MGE357.8415-01.5069 & 17 46 22.8 & --31 33 48 & 28 & 0.04(35) & ? & N & N &&  &  & \\
MGE358.3425-02.4585 & 17 51 25.3 & --31 37 24 & 21 & 0.06(10) & -- & -- & N &&  &  & \\
MGE358.5502+00.7938 & 17 39 01.0 & --29 44 59 & 18 & 0.21(9) & N & N & ? &&  &  & \\
MGE358.8068+00.8872 & 17 39 17.0 & --29 28 59 & 16 & 0.22(6) & N & N & ? &&  &  & \\
MGE358.9577+01.9334 & 17 35 36.5 & --28 47 42 & 24 & 0.16(6) & ? & N & N &&  &  & \\
MGE359.5381-01.0838 & 17 48 46.6 & --29 53 34 & 23 & 0.03(15) & ? & N & ? &&  &  & \\
MGE359.5605-01.3330 & 17 49 49.0 & --30 00 06 & 25 & 0.18(7) & ? & ? & ? &&  &  & \\
MGE359.9600+02.3633 & 17 36 24.6 & --27 43 10 & 29 & 0.03(19) & -- & -- & ? &&  &  & \\
\cutinhead{2c: Rings---bilaterally symmetric}
MGE002.8866-02.7510 & 18 03 02.8 & --27 50 17 & 22 & 0.05(15) & -- & -- & ? &&  &  & \\
MGE005.1020+01.7969 & 17 50 30.6 & --23 38 24 & 18 & 0.16(7) & -- & -- & N &&  &  & \\
MGE013.5944+00.2139 & 18 14 17.1 & --17 02 15 & 24 & 0.39(11) & N & N & N &&  &  & \\
MGE022.5082-00.0086 & 18 32 17.3 & --09 16 13 & 19 & 0.17(31) & ? & N & N &&  &  & \\
MGE029.0784+00.4545 & 18 42 46.8 & --03 13 17 & 28 & 2.1(11) & Y & Y & Y && 2.0 & PN A55 36 & PN\\
MGE032.9141+00.2088 & 18 50 39.7 & +00 04 46 & 30 & 2.2(8) & Y & N & Y && 9.9 & IRAS 18481+0001 & IR\\
MGE033.2929-00.2805 & 18 53 05.7 & +00 11 36 & 25 & 0.28(19) & Y & N & ? &&  &  & \\
MGE044.5870+00.7927 & 19 10 04.3 & +10 43 28 & 27 & 0.16(11) & N & N & N &&  &  & \\
\cutinhead{3a: Disks---flat}
MGE000.1134-00.0975 & 17 46 16.2 & --28 53 24 & 16 & 1.2(13) & ? & N & ? &&  &  & \\
MGE000.6925-01.4687 & 17 53 00.6 & --29 05 52 & 18 & 0.08(7) & N & N & N && 9.6 & JaSt 88 & PN\\
MGE000.7067-01.5718 & 17 53 27.0 & --29 08 17 & 15 & 0.30(3) & ? & N & ? && 1.0 & JaSt2 11 & PN\\
MGE001.0178-01.9642 & 17 55 43.1 & --29 04 04 & 28 & 1.4(2) & ? & Y & Y && 4.5 & PN K 6-35\tablenotemark{b} & PN\\
MGE001.0930+01.4875 & 17 42 28.2 & --27 13 34 & 23 & 0.11(9) & N & N & N && 4.1 & JaSt2 4 & PN\\
MGE001.7985+02.2079 & 17 41 23.5 & --26 14 50 & 20 & 0.03(11) & -- & -- & N &&  &  & \\
MGE001.8248-00.9674 & 17 53 39.5 & --27 52 05 & 16 & 0.12(8) & N & N & ? &&  &  & \\
MGE001.8960-02.5094 & 17 59 52.6 & --28 34 47 & 16 & 0.09(7) & -- & -- & N && 1.2 & PN G001.9-02.5 & PN?\\
MGE002.0411-02.1635 & 17 58 50.3 & --28 16 55 & 20 & 0.06(12) & -- & -- & ? &&  &  & \\
MGE002.1530-01.0607 & 17 54 46.3 & --27 37 56 & 16 & 0.10(10) & ? & N & N &&  &  & \\
MGE002.3535+01.9852 & 17 43 32.1 & --25 53 33 & 22 & 0.04(19) & -- & -- & N &&  &  & \\
MGE002.6242+00.7861 & 17 48 43.4 & --26 17 04 & 12 & 0.02(17) & N & N & -- &&  &  & \\
MGE003.5004+01.8442 & 17 46 42.8 & --24 59 16 & 18 & 0.07(17) & -- & -- & N &&  &  & \\
MGE005.5423-02.8069 & 18 09 02.7 & --25 32 53 & 20 & 0.29(3) & -- & -- & Y &&  &  & \\
MGE005.9143+02.6666 & 17 49 04.3 & --22 29 50 & 24 & 0.02(33) & -- & -- & N &&  &  & \\
MGE006.0270+02.6063 & 17 49 32.8 & --22 25 53 & 18 & 0.03(15) & -- & -- & ? &&  &  & \\
MGE006.4993+01.8661 & 17 53 20.6 & --22 24 10 & 18 & 0.05(18) & -- & -- & N &&  &  & \\
MGE006.7488-01.9183 & 18 08 11.2 & --24 03 47 & 18 & 0.27(14) & -- & -- & ? &&  &  & \\
MGE007.6440+02.0742 & 17 55 03.6 & --21 18 38 & 16 & 0.12(4) & -- & -- & N && 0.2 & PHR J1755-2118 & PN\\
MGE009.4257-01.2294 & 18 11 10.6 & --21 23 15 & 20 & 1.6(2) & -- & -- & -- && 1.1 & PN G009.4-01.2\tablenotemark{b} & PN\\
MGE010.6846-00.6280 & 18 11 30.8 & --19 59 41 & 15 & 0.40(9) & N & N & ? &&  &  & \\
MGE014.1176+00.0816 & 18 15 48.9 & --16 38 27 & 15 & 0.20(46) & ? & N & N &&  &  & \\
MGE015.5409+00.8084 & 18 15 58.5 & --15 02 36 & 20 & 0.06(29) & N & N & ? &&  &  & \\
MGE017.4818+00.8837 & 18 19 29.9 & --13 17 56 & 27 & 0.74(11) & N & N & ? &&  &  & \\
MGE019.6492+00.7740 & 18 24 04.0 & --11 26 16 & 18 & 0.29(4) & ? & ? & Y &&  &  & \\
MGE021.1662+00.9358 & 18 26 22.1 & --10 01 15 & 20 & 0.02(51) & N & N & ? &&  &  & \\
MGE028.8843+00.1226 & 18 43 36.5 & --03 32 44 & 16 & 0.09(36) & N & N & N &&  &  & \\
MGE029.4034-00.4496 & 18 46 35.9 & --03 20 43 & 16 & 0.08(27) & N & N & ? &&  &  & \\
MGE030.1503+00.1237 & 18 45 55.2 & --02 25 08 & 26 & 9.3($<1$) & Y & Y\tablenotemark{k} & Y && 2.2 & IRAS 18433-0228 & IR\\
MGE030.5495+00.9160 & 18 43 49.7 & --01 42 07 & 15 & 0.26(31) & N & N & ? &&  &  & \\
MGE033.4364+00.6435 & 18 50 04.0 & +00 44 33 & 21 & 0.03(54) & N & N & N &&  &  & \\
MGE040.5176-00.1423 & 19 05 50.8 & +06 40 54 & 18 & 0.17(20) & ? & N & ? &&  &  & \\
MGE051.0214-00.4885 & 19 27 06.7 & +15 48 35 & 18 & 0.46(3) & N & Y & Y &&  &  & \\
MGE303.4121-00.8953 & 12 55 46.9 & --63 45 47 & 18 & 0.12(36) & N & N & ? &&  &  & \\
MGE314.5620+00.1984 & 14 28 01.2 & --60 28 26 & 16 & 0.36(3) & N & N & N &&  &  & \\
MGE318.6864+00.1018 & 14 58 02.9 & --58 50 31 & 21 & 0.23(9) & N & N & ? &&  &  & \\
MGE327.7248+01.0008 & 15 48 41.6 & --53 07 04 & 21 & 0.39(8) & N & Y & -- &&  &  & \\
MGE328.2972-00.7801 & 15 59 18.9 & --54 07 39 & 24 & 0.09(17) & N & N & ? &&  &  & \\
MGE329.7690+00.5262 & 16 01 03.7 & --52 10 37 & 30 & 0.27(30) & ? & N & ? &&  &  & \\
MGE329.8273-00.5329 & 16 05 57.6 & --52 55 58 & 16 & 0.53(2) & N & N & Y &&  &  & \\
MGE331.3471-00.5530 & 16 13 22.9 & --51 55 05 & 15 & 0.29(23) & Y & N & ? &&  &  & \\
MGE347.2495-00.3759 & 17 12 53.3 & --39 46 22 & 16 & 0.24(11) & N & N & ? &&  &  & \\
MGE352.7640+01.6844 & 17 20 28.4 & --34 05 05 & 15 & 0.06(13) & -- & -- & N &&  &  & \\
MGE354.0020-01.8419 & 17 38 04.1 & --35 00 08 & 18 & 0.08(8) & -- & -- & -- &&  &  & \\
MGE354.1575-00.3221 & 17 32 16.1 & --34 03 00 & 18 & 0.05(22) & N & N & -- &&  &  & \\
MGE354.7174+02.8360 & 17 21 15.6 & --31 49 28 & 18 & 0.07(13) & -- & -- & ? &&  &  & \\
MGE355.5334-02.1243 & 17 43 09.6 & --33 51 12 & 22 & 0.05(14) & -- & -- & ? &&  &  & \\
MGE355.6162-02.3903 & 17 44 27.7 & --33 55 19 & 33 & 0.11(26) & -- & -- & N && 3.8 & PHR J1744-3355 & PN\\
MGE356.2849+00.9344 & 17 32 47.2 & --31 34 54 & 18 & 0.43(11) & ? & N & -- &&  &  & \\
MGE356.7274-02.6316 & 17 48 12.7 & --33 05 52 & 18 & 0.01(39) & -- & -- & N &&  &  & \\
MGE357.2028-02.2696 & 17 47 54.0 & --32 30 15 & 18 & 0.08(16) & -- & -- & N &&  &  & \\
MGE357.6548+01.0694 & 17 35 43.4 & --30 21 28 & 18 & 0.52(10) & N & N & Y && 2.2 & PK 357+01 3 & PN\\
MGE358.7080+02.4704 & 17 32 55.9 & --28 42 50 & 15 & 0.03(15) & -- & -- & N &&  &  & \\
MGE358.7131+02.6027 & 17 32 26.3 & --28 38 15 & 18 & 0.03(12) & -- & -- & N &&  &  & \\
MGE358.8655+01.2862 & 17 37 52.7 & --29 13 13 & 15 & 0.15(6) & N & N & N &&  &  & \\
MGE359.2412+02.3353 & 17 34 45.8 & --28 20 22 & 24 & 0.09(13) & -- & -- & N &&  &  & \\
MGE359.7169-00.7867 & 17 48 01.8 & --29 35 11 & 16 & 0.17(30) & ? & ? & ? &&  &  & \\
MGE359.7280-00.7201 & 17 47 47.6 & --29 32 33 & 20 & 0.09(30) & N & N & ? &&  &  & \\
MGE359.8429-01.2646 & 17 50 12.8 & --29 43 27 & 20 & 0.13(10) & ? & N & ? &&  &  & \\
\cutinhead{3b: Disks---peaked}
MGE000.1495-01.0708 & 17 50 10.3 & --29 21 42 & 14 & 0.15(7) & Y & N & N &&  &  & \\
MGE000.4284+00.2292 & 17 45 44.7 & --28 27 04 & 14 & 0.08(38) & N & N & -- &&  &  & \\
MGE000.8076+01.1132 & 17 43 13.7 & --27 39 56 & 14 & 0.03(18) & N & N & ? &&  &  & \\
MGE001.6982+00.1362 & 17 49 04.9 & --27 24 47 & 18 & 0.08(16) & N & N & -- &&  &  & \\
MGE002.0287+01.5161 & 17 44 33.5 & --26 24 53 & 14 & 0.04(11) & N & N & ? && 7.8 & JaSt 50 & PN\\
MGE003.0836+01.6435 & 17 46 31.0 & --25 26 53 & 13 & 0.13(5) & -- & -- & ? &&  &  & \\
MGE003.2861+01.3814 & 17 47 58.5 & --25 24 38 & 16 & 0.36(2) & N & N & Y &&  &  & \\
MGE003.7287-00.3847 & 17 55 43.1 & --25 55 51 & 12 & 0.02(22) & N & N & -- &&  &  & \\
MGE004.3111+01.8455 & 17 48 33.1 & --24 17 36 & 18 & 0.33(4) & -- & -- & Y && 0.2 & PHR J1748-2417 & PN\\
MGE004.7474+00.8115 & 17 53 26.0 & --24 26 49 & 13 & 0.03(13) & N & N & -- &&  &  & \\
MGE006.1745-02.1048 & 18 07 40.9 & --24 39 20 & 16 & 0.11(8) & -- & -- & ? && 2.5 & PN G006.1-02.1 & PN\\
MGE006.4395+00.1299 & 17 59 43.8 & --23 19 35 & 16 & 1.2(2) & N & Y & ? &&  &  & \\
MGE006.9367+00.0497 & 18 01 06.4 & --22 56 05 & 20 & 0.10(35) & N & N & N &&  &  & \\
MGE007.1627-00.2781 & 18 02 49.7 & --22 54 01 & 11 & 0.02(71) & ? & N & -- &&  &  & \\
MGE029.2228+00.5392 & 18 42 44.6 & --03 03 15 & 16 & 0.17(16) & N & N & N &&  &  & \\
MGE032.4982+00.1615 & 18 50 04.3 & --00 18 44 & 16 & 0.60(3) & Y & Y & ? && 2.3 & G032.498+0.161\tablenotemark{n} & Rad\\
MGE032.8593+00.2806 & 18 50 18.4 & +00 03 48 & 15 & 0.44(6) & N & N & ? && 1.4 & G032.859+0.280\tablenotemark{n} & Rad\\
MGE040.3704-00.4750 & 19 06 45.8 & +06 23 53 & 27 & 1.1(4) & N & Y\tablenotemark{k} & Y && 2.1 & PN A55 41 & PN\\
MGE044.9328-00.0101 & 19 13 37.2 & +10 39 34 & 21 & 0.07(11) & N & N & ? && 4.4 & PN G044.9+00.0 & PN\\
MGE301.8797-00.1036 & 12 42 10.7 & --62 57 23 & 16 & 0.03(20) & N & N & N &&  &  & \\
MGE352.3117-00.9711 & 17 29 58.3 & --35 56 56 & 18 & 0.37(9) & N & N & Y &&  &  & \\
MGE353.4200-01.0326 & 17 33 12.4 & --35 03 20 & 15 & 0.21(6) & ? & N & -- &&  &  & \\
MGE353.4568+00.8258 & 17 25 47.0 & --33 59 55 & 16 & 0.17(18) & N & N & -- &&  &  & \\
MGE354.5377+02.4713 & 17 22 11.7 & --32 10 45 & 27 & 0.67(2) & -- & -- & Y && 0.5 & PHR J1722-3210\tablenotemark{b} & PN\\
MGE355.0574+02.8195 & 17 22 13.7 & --31 33 15 & 16 & 0.04(14) & -- & -- & ? &&  &  & \\
MGE355.2534+02.0154 & 17 25 52.3 & --31 50 40 & 15 & 0.13(58) & -- & -- & N &&  &  & \\
MGE356.5561-00.7652 & 17 40 13.6 & --32 16 04 & 14 & 0.15(20) & N & N & -- &&  &  & \\
MGE356.6214+01.9982 & 17 29 29.5 & --30 43 01 & 16 & 0.10(8) & -- & -- & N &&  &  & \\
MGE356.8155-00.3843 & 17 39 21.3 & --31 50 44 & 15 & 0.72(4) & ? & N & -- && 5.0 & GPSR5 356.816-0.385 & Rad\\
MGE357.2013+01.5550 & 17 32 41.4 & --30 28 31 & 16 & 0.05(9) & ? & N & N &&  &  & \\
MGE357.7075-00.7184 & 17 42 54.1 & --31 15 56 & 14 & 0.20(16) & N & N & -- &&  &  & \\
MGE357.7077+01.5812 & 17 33 51.9 & --30 02 10 & 16 & 0.07(12) & N & N & N &&  &  & \\
MGE357.7758+02.3326 & 17 31 07.9 & --29 34 10 & 16 & 0.04(24) & -- & -- & N &&  &  & \\
MGE359.3684+01.3182 & 17 38 59.0 & --28 46 41 & 16 & 0.24(4) & N & N & N && 5.1 & JaSt 12 & PN\\
MGE359.6760+01.2410 & 17 40 01.6 & --28 33 32 & 15 & 0.22(5) & ? & N & ? &&  &  & \\
\cutinhead{3c: Disks---bilaterally symmetric}
MGE000.0689-02.1037 & 17 54 04.5 & --29 57 25 & 18 & 0.25(6) & Y & N & ? &&  &  & \\
MGE000.1107+02.0286 & 17 38 03.2 & --27 46 18 & 18 & 0.47(2) & ? & ? & Y && 4.3 & PHR J1738-2746 & PN\\
MGE000.6713+01.4999 & 17 41 25.2 & --27 34 40 & 21 & 0.15(5) & N & N & ? &&  &  & \\
MGE001.0040-01.2822 & 17 52 59.9 & --28 44 06 & 18 & 0.25(5) & N & N & N && 11.5 & JaSt 87 & PN\\
MGE001.0098-02.0666 & 17 56 06.3 & --29 07 34 & 18 & 0.08(15) & N & N & ? &&  &  & \\
MGE002.0599-01.0642 & 17 54 34.4 & --27 42 51 & 27 & 0.95(4) & ? & N & ? &&  &  & \\
MGE002.2728-00.9131 & 17 54 28.2 & --27 27 15 & 18 & 0.35(5) & N & N & N &&  &  & \\
MGE003.0631-01.4329 & 17 58 16.7 & --27 01 56 & 15 & 0.01(43) & N & N & N &&  &  & \\
MGE003.3094+01.0342 & 17 49 20.8 & --25 34 10 & 15 & 0.03(23) & N & N & N &&  &  & \\
MGE003.5533-02.4421 & 18 03 18.4 & --27 06 22 & 30 & 6.0(2) & -- & -- & -- && 1.0 & IC 4673 & PN\\
MGE003.8301-00.1385 & 17 55 00.0 & --25 43 09 & 20 & 7.5(2) & Y & Y & -- && 19.6 & IRAS 17519-2542\tablenotemark{j} & IR\\
MGE006.1437-00.7853 & 18 02 33.7 & --24 02 13 & 18 & 0.06(35) & N & N & -- &&  &  & \\
MGE006.5850-00.0135 & 18 00 35.2 & --23 16 17 & 12 & 0.02(61) & N & N & ? &&  &  & \\
MGE006.7840+01.1298 & 17 56 43.0 & --22 31 42 & 17 & 0.01(62) & ? & N & N &&  &  & \\
MGE007.0582-02.3635 & 18 10 32.9 & --24 00 27 & 18 & 0.20(14) & -- & -- & ? &&  &  & \\
MGE007.7506-00.3392 & 18 04 18.8 & --22 25 07 & 18 & 0.47(6) & Y & ? & ? &&  &  & \\
MGE007.8745+00.7500 & 18 00 29.3 & --21 46 23 & 18 & 0.14(9) & Y & N & ? &&  &  & \\
MGE008.5984-01.1993 & 18 09 20.7 & --22 05 50 & 18 & 0.05(17) & -- & -- & N &&  &  & \\
MGE032.9266-00.7652 & 18 54 09.1 & --00 21 13 & 18 & 0.02(22) & N & N & N &&  &  & \\
MGE034.8961+00.3018 & 18 53 56.8 & +01 53 08 & 21 & 0.27(12) & N & N & N &&  &  & \\
MGE037.9742-00.7954 & 19 03 29.4 & +04 07 20 & 20 & 0.12(14) & ? & N & ? &&  &  & \\
MGE045.9426-00.7527 & 19 18 12.7 & +11 12 26 & 28 & 0.06(24) & ? & N & ? &&  &  & \\
MGE321.4940-00.9818 & 15 20 51.9 & --58 21 43 & 18 & 0.08(10) & Y & ? & N &&  &  & \\
MGE325.5783+00.8074 & 15 37 55.2 & --54 34 25 & 15 & 0.06(7) & N & N & N &&  &  & \\
MGE326.9346+00.9894 & 15 44 34.1 & --53 36 48 & 14 & 0.06(31) & N & N & -- &&  &  & \\
MGE344.1648+00.2733 & 17 00 20.0 & --41 50 43 & 24 & 0.99(3) & N & Y & Y &&  &  & \\
MGE345.4797+00.1407 & 17 05 10.4 & --40 53 06 & 21 & 12.(4) & N & Y & Y && 2.8 & IC 4637 & PN\\
MGE356.4594+01.0730 & 17 32 41.3 & --31 21 35 & 18 & 0.08(16) & ? & N & -- &&  &  & \\
MGE356.5247+02.2864 & 17 28 07.6 & --30 38 17 & 20 & 0.45(6) & -- & -- & Y && 2.4 & PHR J1728-3038 & PN\\
MGE356.8316+01.8709 & 17 30 31.4 & --30 36 42 & 17 & 0.07(82) & -- & -- & N &&  &  & \\
MGE357.3340+01.3123 & 17 33 58.3 & --30 29 45 & 21 & 0.20(6) & N & ? & N && 0.4 & PHR J1733-3029 & PN\\
MGE357.4403+01.5640 & 17 33 15.5 & --30 16 12 & 20 & 0.04(23) & ? & N & N &&  &  & \\
MGE358.0310-01.9836 & 17 48 45.4 & --31 38 51 & 18 & 0.10(26) & Y & N & N &&  &  & \\
MGE358.4433-00.8992 & 17 45 25.0 & --30 44 00 & 15 & 0.07(20) & ? & ? & N &&  &  & \\
MGE358.6888+00.1824 & 17 41 44.8 & --29 57 25 & 10 & 0.06(99) & N & N & -- &&  &  & \\
MGE359.6256-00.9322 & 17 48 23.2 & --29 44 23 & 26 & 0.07(19) & ? & N & ? &&  &  & \\
\cutinhead{3d: Disks---oblong}
MGE000.3521+01.7064 & 17 39 52.0 & --27 44 20 & 16 & 0.18(3) & N & N & Y && 13.4 & JaSt 21 & PN\\
MGE001.2693+00.7174 & 17 45 50.4 & --27 28 44 & 16 & 0.44(2) & Y & N & ? &&  &  & \\
MGE001.5182-02.4587 & 17 58 49.2 & --28 52 55 & 16 & 0.03(10) & -- & -- & ? && 0.6 & PN G001.5-02.4 & PN\\
MGE001.8562-00.4996 & 17 51 54.6 & --27 36 11 & 12 & 0.07(11) & N & N & -- &&  &  & \\
MGE002.1604+01.2401 & 17 45 55.0 & --26 26 46 & 16 & 0.28(3) & N & N & Y && 3.8 & PN G002.1+01.2 & PN?\\
MGE003.2610+01.8522 & 17 46 08.1 & --25 11 17 & 15 & 0.07(11) & -- & -- & ? &&  &  & \\
MGE006.0169+01.2081 & 17 54 44.8 & --23 09 06 & 18 & 0.21(7) & -- & -- & Y && 1.2 & PHR J1754-2309 & PN?\\
MGE007.0999+01.9268 & 17 54 25.7 & --21 51 15 & 20 & 0.06(15) & -- & -- & ? &&  &  & \\
MGE007.4345-01.7575 & 18 09 01.3 & --23 23 08 & 16 & 0.10(13) & -- & -- & N &&  &  & \\
MGE007.9853+00.9983 & 17 59 47.8 & --21 33 13 & 16 & 0.28(6) & N & N & Y &&  &  & \\
MGE008.9409+00.2532 & 18 04 36.3 & --21 05 26 & 18 & 0.49(5) & ? & Y & ? && 1.6 & G008.94097+0.25373\tablenotemark{o} & Rad\\
MGE009.3523+00.4733 & 18 04 38.9 & --20 37 27 & 28 & 2.5(2) & N & Y & Y && 0.2 & PN PBOZ 33 & PN?\\
MGE299.1183-00.1370 & 12 18 00.9 & --62 45 40 & 18 & 0.04(21) & N & N & N && 1.0 & PHR J1218-6245 & PN?\\
MGE318.9322+00.6959 & 14 57 35.5 & --58 12 06 & 28 & 1.8(2) & N & Y & Y && 4.0 & PHR J1457-5818\tablenotemark{b} & PN\\
MGE358.2714-00.4486 & 17 43 12.8 & --30 38 38 & 16 & 0.45(12) & N & N & -- &&  &  & \\
MGE358.8567+01.6271 & 17 36 32.4 & --29 02 42 & 20 & 0.14(5) & N & N & N &&  &  & \\
MGE358.9299-01.9684 & 17 50 51.1 & --30 52 07 & 15 & 0.06(9) & N & N & N &&  &  & \\
MGE359.5784-00.8461 & 17 47 56.1 & --29 44 08 & 26 & 0.23(14) & N & Y & ? &&  &  & \\
\cutinhead{3e: Disks---irregular}
MGE000.4900-02.1986 & 17 55 25.9 & --29 38 29 & 25 & 0.09(55) & -- & -- & Y && 6.8 & OGLEII${\dots}$\tablenotemark{c} & EB*\\
MGE000.6829+01.9996 & 17 39 32.3 & --27 18 11 & 15 & 0.00(40) & N & N & N &&  &  & \\
MGE000.7880+02.9357 & 17 36 14.2 & --26 42 51 & 20 & 0.03(28) & -- & -- & N &&  &  & \\
MGE000.9934+02.2747 & 17 39 13.9 & --26 53 39 & 13 & 0.01(29) & -- & -- & N &&  &  & \\
MGE001.1947-01.0735 & 17 52 37.3 & --28 27 53 & 20 & 0.09(21) & N & N & N &&  &  & \\
MGE001.2760-01.3496 & 17 53 53.5 & --28 32 05 & 24 & 0.09(18) & N & N & N &&  &  & \\
MGE001.2846-01.2034 & 17 53 20.3 & --28 27 11 & 21 & 0.32(5) & N & N & N && 6.1 & JaSt 92 & PN\\
MGE001.4688-01.7089 & 17 55 44.8 & --28 32 59 & 20 & 0.04(21) & ? & N & ? &&  &  & \\
MGE001.6301-02.6875 & 17 59 58.9 & --28 53 55 & 18 & 0.12(15) & -- & -- & ? && 26.6 & PHR J1759-2853\tablenotemark{b} & PN\\
MGE001.9965+00.1976 & 17 49 32.1 & --27 07 32 & 12 & 0.03(85) & N & N & -- &&  &  & \\
MGE002.0557+02.1161 & 17 42 20.6 & --26 04 38 & 20 & 0.01(50) & N & N & N &&  &  & \\
MGE002.3014-01.9492 & 17 58 35.0 & --27 56 59 & 18 & 0.07(7) & N & N & N && 3.6 & PHR J1758-2756 & PN\\
MGE002.4094-02.0786 & 17 59 20.1 & --27 55 14 & 13 & 0.00(33) & ? & N & ? &&  &  & \\
MGE002.8530+01.4514 & 17 46 42.9 & --25 44 41 & 18 & 0.07(9) & ? & N & ? &&  &  & \\
MGE003.5014+01.3314 & 17 48 39.3 & --25 15 07 & 20 & 0.03(18) & ? & N & ? &&  &  & \\
MGE004.0775+02.2998 & 17 46 18.9 & --24 15 30 & 24 & 0.05(22) & -- & -- & N &&  &  & \\
MGE004.0816+01.1591 & 17 50 37.4 & --24 50 34 & 26 & 0.50(6) & ? & N & ? &&  &  & \\
MGE004.1668+02.4238 & 17 46 03.2 & --24 07 04 & 24 & 0.08(15) & -- & -- & N &&  &  & \\
MGE004.2374-00.8916 & 17 58 48.0 & --25 44 41 & 18 & 0.18(5) & ? & N & N &&  &  & \\
MGE004.2573-02.9981 & 18 07 01.4 & --26 45 49 & 42 & 0.39(10) & -- & -- & ? &&  &  & \\
MGE004.2590-01.4204 & 18 00 53.3 & --25 59 20 & 23 & 0.06(16) & ? & N & N &&  &  & \\
MGE004.6913+01.8888 & 17 49 14.7 & --23 56 43 & 21 & 0.02(28) & -- & -- & N && 2.3 & PN G004.6+01.8 & PN\\
MGE004.7784+01.2007 & 17 52 01.8 & --24 13 21 & 23 & 0.07(13) & N & N & N &&  &  & \\
MGE004.8004+01.7757 & 17 49 54.9 & --23 54 35 & 25 & 0.10(9) & -- & -- & N &&  &  & \\
MGE005.1516+00.0444 & 17 57 14.4 & --24 29 06 & 20 & 0.24(16) & N & N & -- &&  &  & \\
MGE005.2641+00.3775 & 17 56 13.4 & --24 13 13 & 16 & 1.1(2) & ? & Y & -- && 2.7 & G005.264+0.377\tablenotemark{n} & Rad\\
MGE006.6637-01.1965 & 18 05 14.8 & --23 47 09 & 27 & 0.05(23) & -- & -- & N &&  &  & \\
MGE006.7691-02.7365 & 18 11 22.5 & --24 26 24 & 24 & 0.21(5) & -- & -- & ? &&  &  & \\
MGE007.0571+02.0063 & 17 54 02.4 & --21 51 04 & 15 & 0.01(31) & -- & -- & N &&  &  & \\
MGE007.6985+02.1443 & 17 54 55.1 & --21 13 41 & 24 & 0.02(33) & -- & -- & N &&  &  & \\
MGE008.4067-01.6982 & 18 10 50.0 & --22 30 21 & 16 & 0.12(4) & -- & -- & N && 16.0 & G008.405-1.694\tablenotemark{n} & Rad\\
MGE010.0155-00.6520 & 18 10 13.7 & --20 35 33 & 21 & 0.06(24) & N & N & N &&  &  & \\
MGE012.0367+01.0149 & 18 08 11.6 & --18 01 01 & 32 & 0.09(40) & N & N & N &&  &  & \\
MGE020.8749+00.9698 & 18 25 41.6 & --10 15 46 & 18 & 0.04(21) & N & N & N &&  &  & \\
MGE021.4030-00.4666 & 18 31 51.8 & --10 27 43 & 21 & 0.09(18) & ? & N & ? &&  &  & \\
MGE022.5964+00.9098 & 18 29 09.4 & --08 46 01 & 18 & 0.06(25) & N & N & N &&  &  & \\
MGE027.5373+00.5473 & 18 39 37.3 & --04 32 57 & 15 & 0.07(23) & N & N & N &&  &  & \\
MGE034.9249+00.4834 & 18 53 21.1 & +01 59 39 & 15 & 0.09(38) & N & N & ? &&  &  & \\
MGE035.7017-00.2975 & 18 57 33.1 & +02 19 45 & 19 & 0.06(17) & N & Y & Y &&  &  & \\
MGE042.3030+00.6323 & 19 06 22.9 & +08 37 24 & 28 & 0.04(31) & N & Y & ? &&  &  & \\
MGE046.8232-00.2257 & 19 17 59.4 & +12 13 55 & 18 & 0.03(36) & N & N & ? &&  &  & \\
MGE054.6957+00.7074 & 19 30 04.6 & +19 36 31 & 18 & 0.08(13) & N & Y & Y &&  &  & \\
MGE059.2301+00.4005 & 19 40 43.1 & +23 25 15 & 21 & 0.07(12) & N & N & N &&  &  & \\
MGE065.9141+00.5966 & 19 55 02.4 & +29 17 20 & 33 & 1.1(4) & -- & -- & Y && 0.8 & NGC 6842 & PN\\
MGE303.9339-00.3946 & 13 00 20.6 & --63 14 56 & 25 & 0.10(20) & ? & Y & Y &&  &  & \\
MGE318.5117-00.5950 & 14 59 23.8 & --59 32 21 & 28 & 0.29(10) & ? & N & ? &&  &  & \\
MGE318.6228+00.7522 & 14 55 18.5 & --58 17 41 & 15 & 0.01(31) & N & N & N &&  &  & \\
MGE328.1916-00.7829 & 15 58 46.7 & --54 11 54 & 24 & 1.1(10) & ? & ? & Y &&  &  & \\
MGE334.3288-00.7978 & 16 27 55.3 & --49 59 30 & 40 & 0.40(30) & Y & Y & Y &&  &  & \\
MGE353.9103+00.4858 & 17 28 21.7 & --33 48 41 & 15 & 0.04(47) & ? & N & N &&  &  & \\
MGE355.2478+00.9935 & 17 29 52.3 & --32 24 59 & 18 & 0.16(9) & ? & N & -- &&  &  & \\
MGE355.4167+01.0342 & 17 30 09.1 & --32 15 11 & 14 & 0.01(87) & ? & N & -- &&  &  & \\
MGE355.5953-02.1285 & 17 43 20.0 & --33 48 11 & 15 & 0.02(29) & -- & -- & N &&  &  & \\
MGE355.8086+01.2344 & 17 30 22.8 & --31 48 57 & 15 & 0.07(14) & ? & N & -- &&  &  & \\
MGE355.8839-01.4270 & 17 41 12.1 & --33 11 18 & 18 & 0.19(4) & ? & N & Y &&  &  & \\
MGE356.0998+00.7001 & 17 33 14.2 & --31 51 52 & 18 & 0.06(18) & ? & N & N &&  &  & \\
MGE356.1447+00.0550 & 17 35 54.5 & --32 10 35 & 25 & 0.44(19) & Y & N & ? && 8.9 & GPSR5 356.144+0.053 & Rad\\
MGE356.4537+00.8358 & 17 33 36.4 & --31 29 37 & 15 & 0.12(12) & N & ? & N &&  &  & \\
MGE356.5084+01.0838 & 17 32 46.3 & --31 18 46 & 24 & 0.32(5) & N & N & -- &&  &  & \\
MGE356.6408-00.5704 & 17 39 39.5 & --32 05 33 & 15 & 0.24(12) & N & N & -- &&  &  & \\
MGE356.7168-01.7246 & 17 44 29.6 & --32 38 11 & 18 & 0.60(3) & Y & N & Y && 12.8 & IRAS 17412-3236 & IR\\
MGE356.7415+00.7421 & 17 34 42.4 & --31 18 10 & 16 & 0.23(43) & N & N & N &&  &  & \\
MGE357.8187+00.2875 & 17 39 11.7 & --30 38 19 & 16 & 0.20(46) & N & N & -- &&  &  & \\
MGE357.9893+02.8341 & 17 29 44.7 & --29 06 57 & 21 & 0.21(4) & -- & -- & ? &&  &  & \\
MGE358.0888-01.5465 & 17 47 08.4 & --31 22 21 & 20 & 0.08(20) & N & N & N &&  &  & \\
MGE358.1166+01.0764 & 17 36 50.8 & --29 57 53 & 15 & 0.03(31) & N & N & N &&  &  & \\
MGE358.1558+00.7594 & 17 38 10.8 & --30 06 06 & 24 & 0.22(8) & N & N & -- &&  &  & \\
MGE358.2600+01.4563 & 17 35 43.6 & --29 38 22 & 19 & 0.05(18) & N & N & N &&  &  & \\
MGE358.5717+02.5207 & 17 32 24.1 & --28 48 02 & 26 & 0.05(19) & -- & -- & N &&  &  & \\
MGE358.5824+02.6873 & 17 31 47.4 & --28 42 03 & 33 & 2.2(1) & -- & -- & Y && 1.5 & PK 358+02 5\tablenotemark{b} & PN\\
MGE358.6633-01.3135 & 17 47 35.8 & --30 45 38 & 15 & 0.01(22) & N & N & N &&  &  & \\
MGE358.9738-02.2830 & 17 52 13.1 & --30 59 28 & 21 & 0.02(23) & -- & -- & ? &&  &  & \\
MGE359.1142-01.6065 & 17 49 50.6 & --30 31 30 & 20 & 0.02(45) & ? & N & ? &&  &  & \\
MGE359.4290-00.5034 & 17 46 13.7 & --29 41 09 & 15 & 0.16(26) & ? & N & -- &&  &  & \\
MGE359.4372-02.6997 & 17 54 59.1 & --30 48 09 & 18 & 0.02(21) & N & N & ? &&  &  & \\
MGE359.5136-02.6056 & 17 54 47.1 & --30 41 21 & 35 & 0.02(34) & ? & N & ? &&  &  & \\
MGE359.6095-01.9211 & 17 52 16.2 & --30 15 36 & 29 & 0.03(46) & N & ? & ? &&  &  & \\
MGE359.8693+01.0316 & 17 41 18.1 & --28 30 21 & 36 & 0.04(49) & N & N & ? &&  &  & \\
\cutinhead{4: Two-lobed}
MGE000.7126+02.7811 & 17 36 38.4 & --26 51 39 & 19 & 0.02(21) & -- & -- & N &&  &  & \\
MGE000.9066+02.1503 & 17 39 29.9 & --27 02 01 & 18 & 0.20(3) & -- & -- & ? &&  &  & \\
MGE001.0651+02.7431 & 17 37 37.7 & --26 35 01 & 18 & 0.06(7) & -- & -- & -- &&  &  & \\
MGE001.0839+01.9163 & 17 40 48.7 & --27 00 27 & 18 & 0.05(14) & N & N & N &&  &  & \\
MGE003.4305-01.0738 & 17 57 42.5 & --26 32 05 & 19 & 0.05(16) & N & N & N &&  &  & \\
MGE003.8212+02.6722 & 17 44 20.1 & --24 17 00 & 18 & 0.05(13) & -- & -- & N &&  &  & \\
MGE004.0396-02.7139 & 18 05 26.2 & --26 48 56 & 23 & 0.07(14) & -- & -- & -- && 7.1 & PN G004.0-02.7 & PN\\
MGE007.3587+01.7685 & 17 55 34.7 & --21 42 39 & 16 & 0.36(2) & -- & -- & ? && 0.9 & PN G007.3+01.7 & PN\\
MGE007.6093+02.0988 & 17 54 53.7 & --21 19 41 & 30 & 0.06(24) & -- & -- & N &&  &  & \\
MGE008.1733+01.0914 & 17 59 51.1 & --21 20 39 & 18 & 0.12(15) & N & N & N &&  &  & \\
MGE008.8168+02.2819 & 17 56 48.7 & --20 11 34 & 24 & 0.05(31) & -- & -- & -- &&  &  & \\
MGE016.1274+00.3327 & 18 18 51.6 & --14 45 10 & 18 & 0.31(14) & N & N & ? &&  &  & \\
MGE019.8603-00.3579 & 18 28 33.4 & --11 46 43 & 31 & 1.2(9) & Y & ? & ? && 11.6 & IRAS 18257-1148 & IR\\
MGE030.8780+00.6993 & 18 45 12.0 & --01 30 31 & 18 & 0.15(13) & N & N & N &&  &  & \\
MGE031.3183+00.9958 & 18 44 56.9 & --00 58 54 & 18 & 0.05(21) & N & N & N &&  &  & \\
MGE039.5964+00.4751 & 19 01 56.3 & +06 08 46 & 20 & 0.05(31) & ? & N & N &&  &  & \\
MGE042.7665+00.8222 & 19 06 33.6 & +09 07 20 & 33 & 1.1(1) & Y & Y & Y && 5.2 & IRAS 19041+0902 & IR\\
MGE334.1114+00.3789 & 16 21 47.3 & --49 19 23 & 21 & 0.99(4) & Y & Y\tablenotemark{k} & Y &&  &  & \\
MGE355.7638+00.1424 & 17 34 35.2 & --32 26 58 & 16 & 0.28(10) & ? & N & N && 9.9 & NGC 6383 33 & *iC\\
MGE356.1617-00.5486 & 17 38 21.8 & --32 29 10 & 18 & 0.05(38) & ? & N & N &&  &  & \\
MGE356.5146-01.3137 & 17 42 19.7 & --32 35 35 & 33 & 0.18(16) & ? & N & ? &&  &  & \\
MGE357.2784-02.6349 & 17 49 34.3 & --32 37 38 & 18 & 0.04(30) & -- & -- & ? &&  &  & \\
MGE358.0815+00.8514 & 17 37 38.2 & --30 06 55 & 15 & 0.29(5) & Y & N & N &&  &  & \\
MGE358.2846-00.2593 & 17 42 29.8 & --30 31 59 & 18 & 0.13(20) & ? & N & -- &&  &  & \\
\cutinhead{5: Filamentary}
MGE011.1805-00.3471 & 18 11 28.9 & --19 25 29 & 260 & 33.(18) & ? & N & ? && 22.6 & SNR G011.2-00.3\tablenotemark{b} & SNR\\
MGE027.3891-00.0079 & 18 41 19.8 & --04 56 06 & 250 & 21.(89) & ? & ? & ? && 9.9 & LMH 31\tablenotemark{b} & SNR\\
\cutinhead{6: Miscellaneous}
MGE003.7032-01.7927 & 18 01 06.2 & --26 39 21 & 60 & 0.01(68) & -- & -- & N &&  &  & \\
MGE038.7425-00.6986 & 19 04 33.5 & +04 50 58 & 14 & 0.04(14) & Y & N & N &&  &  & \\
MGE304.6649-00.0439 & 13 06 38.7 & --62 51 52 &  & 0.32(99) & Y & N & N && 2.5 & M01-221 & *\\
MGE305.3881+00.0804 & 13 12 52.5 & --62 41 21 & 90 & 0.13(94) & Y & N & N &&  &  & \\
MGE314.2378+00.9793 & 14 23 21.0 & --59 51 34 & 63 & 0.11(99) & Y & Y & ? &&  &  & \\
MGE316.8732-00.5991 & 14 47 45.1 & --60 17 02 & 15 & 0.04(18) & Y & Y & Y && 7.0 & [JKK2007] G2 & G\\
MGE317.0392-00.4974 & 14 48 36.0 & --60 07 13 & 18 & 0.08(15) & N & Y\tablenotemark{k} & Y && 23.7 & [JKK2007] G1 & G\\
MGE351.2381-00.0145 & 17 23 04.4 & --36 18 20 & 40 & 0.41(28) & N & Y & Y &&  &  & \\
MGE353.5188+01.5107 & 17 23 13.5 & --33 33 44 & 21 & 0.11(13) & -- & -- & -- &&  &  & \\
MGE356.3395+02.0502 & 17 28 33.9 & --30 55 23 & 14 & 0.03(18) & -- & -- & ? &&  &  & \\
\enddata
\tablenotetext{a}{
Although cut off from the MIPSGAL and GLIMPSE coverage, an MSX 8 {\micron} source, 
associated with Wray 15-1680, lies at the center of the ring. IRAS 17173-3305 is 
likely associated with the NE rim, but cannot be confirmed due to lack of coverage}
\tablenotetext{b}{
As mentioned in the text, a number of objects have secondary
identifications not yet associated as such in SIMBAD. These are:\newline
MGE001.0178-01.9642: IRAS 17525-2903\newline
MGE001.2124-02.0748: PHR J1756-2857\newline
MGE001.6301-02.6875: V* V3987 Sgr\newline
MGE003.5282+02.6745: IRAS 17405-2430\newline
MGE009.4257-01.2294: IRAS 18081-2123\newline
MGE011.1805-00.3471: PSR J1811-1926\newline
MGE027.3891-00.0079: PSR J1841-0456\newline
MGE060.4530+00.0712: IRAS 19425+2411\newline
MGE313.3544+00.3123: IRAS 14147-6033\newline
MGE318.9322+00.6959: IRAS 14538-5800\newline
MGE349.7294+00.1747: IRAS 17146-3723\newline
MGE353.5300+02.1882: IRAS 17173-3305\newline
MGE354.5377+02.4713: IRAS 17189-3207\newline
MGE358.5016-00.5048: IRAS 17408-3027\newline
MGE358.5824+02.6873: F3R 18}
\tablenotetext{c}{Full name: OGLEII DIA BUL-SC39 4483}
\tablenotetext{d}{Full name: 2MASS J18393224-0544204}
\tablenotetext{e}{IRAS 18588+0350 is associated with the SE rim}
\tablenotetext{f}{IRAS 16254-7439 is associated with the SE rim}
\tablenotetext{g}{IRAS 16228-5014 is associated with the SW rim}
\tablenotetext{h}{IRAS 16278-4808 is associated with the NW rim}
\tablenotetext{i}{IRAS 15293-5602 may be associated with the SE rim}
\tablenotetext{j}{IRAS 18468-0109 may be associated with the WSW rim}
\tablenotetext{k}{Extended emission seen in all IRAC bands}
\tablenotetext{l}{Extended emission is partially saturated}
\tablenotetext{m}{Central source is saturated}
\tablenotetext{n}{Object matched in White et al. (2005) catalog of compact Galactic radio sources}
\tablenotetext{o}{Object matched in MAGPIS (Helfand et al. 2006) catalog of Galactic radio sources}
\end{deluxetable}

\end{document}